\documentclass[aps,pre,reprint,groupedaddress]{revtex4-1}
\usepackage{amsmath,amssymb}%,mathabx}
\numberwithin{equation}{section}
\usepackage{graphicx,esint,ulem}
\usepackage{mathrsfs}
\usepackage{bm}
\usepackage{enumerate}
\usepackage[english]{babel}
\usepackage[colorlinks=true, urlcolor=blue, citecolor = black]{hyperref}
\usepackage{color}
\usepackage[shortlabels,inline]{enumitem}
\usepackage{tikz}
\usetikzlibrary{calc}
\usetikzlibrary{intersections}
\usetikzlibrary{decorations.pathmorphing}
\usetikzlibrary{decorations.pathreplacing}
\tikzset{wiggly/.style={decorate, decoration={snake,amplitude=2pt,segment length=8pt}}}
\tikzset{brace/.style={decorate,decoration={brace,amplitude=5pt}}}
\newcommand{\R}{\mathbb{R}}

\newcommand{\E}{\mathbb{E}}
\newcommand{\Lgen}{\mathcal{L}}
\newcommand{\D}{\partial}

\providecommand{\eps}{\varepsilon}

\newcommand{\inner}[1]{\left\langle#1\right\rangle}

\newcommand{\diag}{\text{diag}}
\newcommand{\wt}{}%{\widetilde}  % mhc commented out

\newcommand{\bfl}{{\bm l}}
\newcommand{\bff}{{\bm f}}
\newcommand{\bfs}{{\bm s}}

\newcommand{\bpi}{{\bm \pi}}

\newcommand{\Et}{E_{tether}}  % energy of a tether, disc model
\newcommand{\half}{\frac{1}{2}}  % half
\newcommand{\dd}[2]{\frac{d#1}{d#2}}

\definecolor{todoGreen}{rgb}{0.0, 0.5, 0.0}
\begin{document}

% Use the \preprint command to place your local institutional report
% number in the upper righthand corner of the title page in preprint mode.
% Multiple \preprint commands are allowed.
% Use the 'preprintnumbers' class option to override journal defaults
% to display numbers if necessary
%\preprint{}

%Title of paper
\title{Modeling the relative dynamics of DNA-coated colloids}
%\title{Modeling the friction and diffusion induced on DNA-coated colloids}
%\title{Modeling the friction and diffusion of a colloid moving on a DNA-coated plane}

% repeat the \author .. \affiliation  etc. as needed
% \email, \thanks, \homepage, \altaffiliation all apply to the current
% author. Explanatory text should go in the []'s, actual e-mail
% address or url should go in the {}'s for \email and \homepage.
% Please use the appropriate macro foreach each type of information

% \affiliation command applies to all authors since the last
% \affiliation command. The \affiliation command should follow the
% other information
% \affiliation can be followed by \email, \homepage, \thanks as well.
\author{James P. Lee-Thorp}
%\email[]{leethorp@cims.nyu.edu}
%\homepage[]{Your web page}
%\thanks{}
%\altaffiliation{}
\affiliation{Courant Institute of Mathematical Sciences, New York University, NY, USA}

\author{Miranda Holmes-Cerfon}
\email[]{holmes@cims.nyu.edu}
%\homepage[]{Your web page}
%\thanks{}
%\altaffiliation{}
\affiliation{Courant Institute of Mathematical Sciences, New York University, NY, USA}

%Collaboration name if desired (requires use of superscriptaddress
%option in \documentclass). \noaffiliation is required (may also be
%used with the \author command).
%\collaboration can be followed by \email, \homepage, \thanks as well.
%\collaboration{}
%\noaffiliation

\date{\today}

\begin{abstract}
We construct a theoretical model for the dynamics of a microscale colloidal particle, modeled as an interval, moving horizontally on a DNA-coated surface, modelled as a line coated with springs that can stick to the interval. 
Averaging over the fast DNA dynamics leads to an evolution equation for the particle in isolation, which contains both friction and diffusion.
The  DNA-induced friction coefficient depends on the physical properties of the DNA, and substituting parameter values typical of a 1$\mu$m colloid coated densely with weakly interacting DNA 
gives a coefficient about 100 times larger than the corresponding coefficient of hydrodynamic friction. 
We use a mean-field extension of the model to higher dimensions to estimate the friction tensor for a disc rotating and translating horizontally along a line. When the DNA strands are very stiff and short, the friction coefficient for the disc rolling approaches zero while the friction for the disc sliding remains large. 
Together, these results could have significant implications for the dynamics of DNA-coated colloids or other ligand-receptor systems,
implying that DNA-induced friction between colloids can be stronger than hydrodynamic friction and should be incorporated into simulations, and that it depends nontrivially on the type of relative motion, possibly causing the particles to assemble into out-of-equilibrium metastable states governed by the pathways with the least friction.
\end{abstract}

% insert suggested PACS numbers in braces on next line
% insert suggested keywords - APS authors don't need to do this
%\keywords{}

\maketitle

%%%MAIN TEXT%%%%
\section{Introduction\label{intro}}

Particles that live on the nanoscale or microscale (colloids) %form the building blocks of many of the materials we encounter in everyday life, such as paint, ketchup, toothpaste, concrete, etc. Because such particles are easy to synthesize in large quantities, they 
are widely studied, partly because they 
could be designed to make new materials with novel plasmonic, photonic, biomimetic, programmability, or other properties \cite{Whitesides:2002gba,Porter:2017jf,Lieleg:2011fh,Fan:2011hr,Cheng:2006ju,Stimulak:2014ce}. 
One way to achieve highly specific, programmable interactions between particles is to coat them with strands of DNA, such that each DNA strand has one end glued to the particle surface, and the other end, the ``sticky'' end, is single-stranded with some particular sequence of nucleotides. When two DNA-coated colloids with complementary sticky ends approach each other, the sticky ends hybridize, creating an effective attractive interaction between the particles. A good metaphor for the interaction is to imagine tennis balls coated with velcro, which stick together when they are close enough. 

This technique allows for enormous control over particles and their interactions -- for example, highly specific interactions may be created by mixing DNA with different sticky ends on each particle \cite{Macfarlane:2011fh,Wang:2017bd}, directional interactions may be mimicked by placing DNA in isolated patches \cite{Wang:2012gd} or on clusters of particles \cite{Ducrot:2017cs}, and particles may be synthesized to have a wide variety of shapes \cite{Sacanna:2011dd,Sacanna:2013ge}. 
Such a large parameter space of possibilities cannot be exhaustively searched experimentally, so theory and simulation are required to help design particles to assemble into a desired material \cite{Mladek:2012dka}. However, simulations which include the full dynamics of each DNA strand are infeasible, since there could be on the order of $10^3-10^5$ strands on each particle. Therefore, a challenge for theory is to develop simplified, predictive models that take the complex particle-DNA system and map it to a system that is easier to simulate or otherwise analyze. 

To date there have been several studies of the equilibrium behaviour of DNA-coated colloids, which suggest that for large enough particles, a system of DNA-coated colloids can be modelled as particles interacting with a pair potential, whose shape may be obtained from the characteristics of the DNA \cite{Dreyfus:2009gl,Rogers:2011et,Mognetti:2012je,Varilly:2012gla,AngiolettiUberti:2016dd}. However,  there are no models for the relative dynamics of DNA-coated colloids: how the colloids move around relative to each other, once their DNA strands are interacting.  
A small number of studies have considered other kinds of kinetic effects, such as 
 the overall binding/unbinding rates between two particles \cite{Rogers:2013dc}, the aggregation rates of large collection of particles \cite{Wu:2013jg}, the rate at which a particle approaches a surface \cite{Mani:2012di},  the time-dependent limitations on valency induced by mobile DNA and slow DNA kinetics \cite{JanBachmann:2016dm,Parolini:2016ho}, or the diffusive or subdiffusive behaviour of a particle which moves by unbinding, diffusing in free space, and binding again \cite{Xu:2011gs}.
 Most of these studies are aimed at understanding the overall rate of bond formation between the two colloids; none so far has been able to discriminate between different types of relative spatial motion, or to provide a specific estimate for how quickly the particles move relative to each other. 
Indeed, it is only very recently that DNA-coated colloids have been created that \emph{can} move relative to each other %, on reasonable timescales and over reasonable ranges of temperatures, 
while bound  \cite{Wang:2015ep}. 
Such relative motion is critical if the system is to equilibrate or reach an ordered metastable state. 
Yet, because microscale colloids diffuse slowly, they don't necessarily reach equilibrium, so the kinetic pathways they follow as they assemble may govern the structures they form \cite{Jenkins:2014js,JanBachmann:2016dm,Wang:2017bd}. 

\begin{figure}
% color of light circles
\definecolor{lightcol}{rgb}{0.8,0.8,0.8}

\begin{center}

\begin{tikzpicture}[scale=1]
\pgfmathsetmacro{\R}{0.58};   % radius of circles (should be 0.691; don't understand why not)
\pgfmathsetmacro{\rarc}{\R*0.8};   % radius of arcs
\pgfmathsetmacro{\s}{(1+\R)*1.2};   % shift for arcs
\pgfmathsetmacro{\th}{72};   % angle of polygon
\pgfmathsetmacro{\thS}{-150};  % start angle of arcs
\pgfmathsetmacro{\thE}{-80};  % end angle of arcs

% centers of circles
\coordinate (A) at ({-sin(144)},{-cos(36)});
\coordinate (B) at ({sin(144)},{-cos(36)});
\coordinate (C) at ({sin(72)},{cos(72)});
\coordinate (D) at (0,1);
\coordinate (E) at ({-sin(72)},{cos(72)});

% Start of arcs
\coordinate (Aa) at ({\s*-sin(144)},{\s*-cos(36)});
\coordinate (Ba) at ({\s*sin(144)},{\s*-cos(36)});
\coordinate (Ca) at ({\s*sin(72)},{\s*cos(72)});
\coordinate (Da) at (0,\s*1);
\coordinate (Ea) at ({\s*-sin(72)},{\s*cos(72)});

% Draw circles
\draw[black,fill=red] (A) circle (\R);
\draw[black,fill=lightcol] (B) circle (\R);
\draw[black,fill=lightcol] (C) circle (\R);
\draw[black,fill=lightcol] (D) circle (\R);
\draw[black,fill=blue] (E) circle (\R);

% Draw arcs
\draw[->,red] (Aa) arc[radius=\rarc,start angle=\thS, end angle=\thE] ;
\draw[<-] (Ba) arc[radius=\rarc,start angle=\thS+\th, end angle=\thE+\th];
\draw[->] (Ca) arc[radius=\rarc,start angle=\thS+2*\th, end angle=\thE+2*\th];
\draw[<-] (Da) arc[radius=\rarc,start angle=\thS+3*\th, end angle=\thE+3*\th] ;
\draw[->,blue] (Ea) arc[radius=\rarc,start angle=\thS+4*\th, end angle=\thE+4*\th] ;
\end{tikzpicture}
\qquad
\raisebox{0.75cm}{
\begin{tikzpicture}[scale=1]
\pgfmathsetmacro{\R}{0.58};   % radius of circles (should be 0.691; don't understand why not)

% centers of circles
\coordinate (A) at (0,0);
\coordinate (B) at (2*\R,0);
\coordinate (C) at (4*\R,0);
\coordinate (D) at (1*\R,{sqrt(3)*\R});
\coordinate (E) at (3*\R,{sqrt(3)*\R});

% Draw circles
\draw[black,fill=red] (A) circle (\R);
\draw[black,fill=lightcol] (B) circle (\R);
\draw[black,fill=lightcol] (C) circle (\R);
\draw[black,fill=blue] (D) circle (\R);
\draw[black,fill=lightcol] (E) circle (\R);

\end{tikzpicture}
}
\end{center}
\caption{
The structures that DNA-coated colloids form may be governed by the pathways with the lowest friction, as illustrated in this example. Left: if friction for sliding is high, then particles may get stuck for long times in more open arrangements, like this arrangement of discs. No matter how the particles collectively move there is always a pair which must slide against each other (here the shaded blue/red pair), making the system effectively jammed in this state. Right: if friction for sliding is not prohibitively high, particles may rearrange to form more compact structures.}
\label{fig:slideroll}
\end{figure}
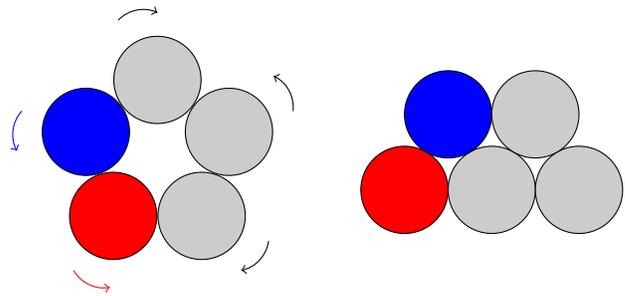

How then will the dynamics of particles interacting via DNA, be modified from those of a standard system of point particles with pair interactions? 
One might expect the DNA to introduce an additional friction between particles, and to correspondingly modify their relative diffusion rate through its fluctuations. If so, the friction and diffusion could depend non-trivially on the type of relative motion. To see why, return to the metaphor of velcro-coated particles, and imagine two pieces of velcro stuck together. If you pull them tangentially to each other, they typically stay stuck -- you have to apply a very large force to get the velcro surfaces to slide horizontally. Therefore, we might expect friction for particles to slide on each others' surfaces to be large.   However, if you take one end of the velcro and pull it upwards and horizontally at the same time, the velcro comes off easily; by applying this motion to a long line of velcro you can easily make the contact patch move down the line. 
Such a motion is analogous to particles rolling on each others' surfaces, so we might expect the friction for rolling to be small. If the rolling friction is much smaller than the sliding friction, this could have a significant impact on the pathways the system is likely to follow, for example promoting more open structures with fewer contacts, as illustrated in Figure \ref{fig:slideroll}. 

Of course, the velcro metaphor could be misleading, since it does not account for the fluctuations of DNA binding and unbinding and changing its length in solution. A different thought experiment suggests that if DNA binds and unbinds rapidly enough, then it never binds for long enough to exert any significant force, so the frictions for different types of relative motions could be comparable, and small. 

Which of these thought experiments is closer to the truth, if any? We aim to answer this question by constructing a coarse-grained model for the relative dynamics of DNA-coated particles. 
Our approach is to 
build a microscopic model for the joint dynamics of particles and DNA, and then average over the fast DNA fluctuations to obtain an equation for the particles in isolation. 
The model and result will be valid for particles whose radius is much larger than a typical DNA strand, so that DNA-induced multiparticle interactions may be neglected, and when the DNA density is large enough and interactions weak enough that fluctuations in coverage are negligible. 

We begin by considering the simplest possible model that gives rise to nontrivial coarse-grained dynamics: a one-dimensional interval moving on a one-dimensional line coated with sticky spring-like tethers (Section \ref{sec:1d}). 
This model gives pedagogical insight into the origin the DNA-induced friction and stochastic forcing with a minimum of parameters, and is amenable to rigorous analysis. 
We perform formal homogenization to derive the coarse-grained dynamics, which include both a DNA-induced friction term, and a related DNA-induced stochastic forcing which causes diffusion. 
We obtain an analytic expression for the friction coefficient (Section \ref{sec:effective}), as a function of the physical parameters of the DNA. 
%The scaling ansatz is in section \ref{sec:ansatz}, the main result is in section \ref{sec:effective}, and the derivation is in section \ref{sec:derivation}. 

We use this analytic expression to estimate the magnitude of the friction coefficient in a typical system where colloids can rearrange on each others' surfaces (Section \ref{sec:physics}), and predict that DNA-induced friction is about 100 times larger than hydrodynamic friction. 
This prediction could be tested experimentally, for example by measuring the diffusion coefficient of a DNA-coated colloid on a DNA-coated plane.

%While this one-dimensional model is modest compared to our ultimate goal of constructing a fully three-dimensional model with complex geometry, 
Our model generalizes via mean-field arguments to higher dimensions, which gives a formula for comparing frictions for different directions of motion (but not their absolute magnitudes.) %(Section \ref{sec:effective}). 
We estimate the effective friction tensor for a disc moving along a DNA-coated line in two dimensions (Section \ref{sec:2d}). The model predicts that when the DNA is short and stiff, as it is in typical systems, then friction associated with the disc rolling approaches zero while the friction associated with sliding remains comparatively large. 

Together, our results suggest that to correctly model the self-assembly of DNA-coated particles, it is critical to include the coarse-grained effect of DNA on the dynamics, and not just on the thermodynamics, and that this effect may sometimes be more important to incorporate than hydrodynamics. We speculate in the conclusion (Section \ref{sec:conclusion}) on the consequences of these predictions, as well as on how this model could be experimentally verified and further developed into a fully three-dimensional model with complex geometry. 
%These results also justify future efforts to build a more realistic, higher-dimensional model of DNA-coated particles that includes more complex geometry and dynamics. 

\section{One-dimensional model}\label{sec:1d}

\subsection{Overview}\label{sec:overview1d}

\begin{figure}
\centering
\includegraphics[width=1\columnwidth]{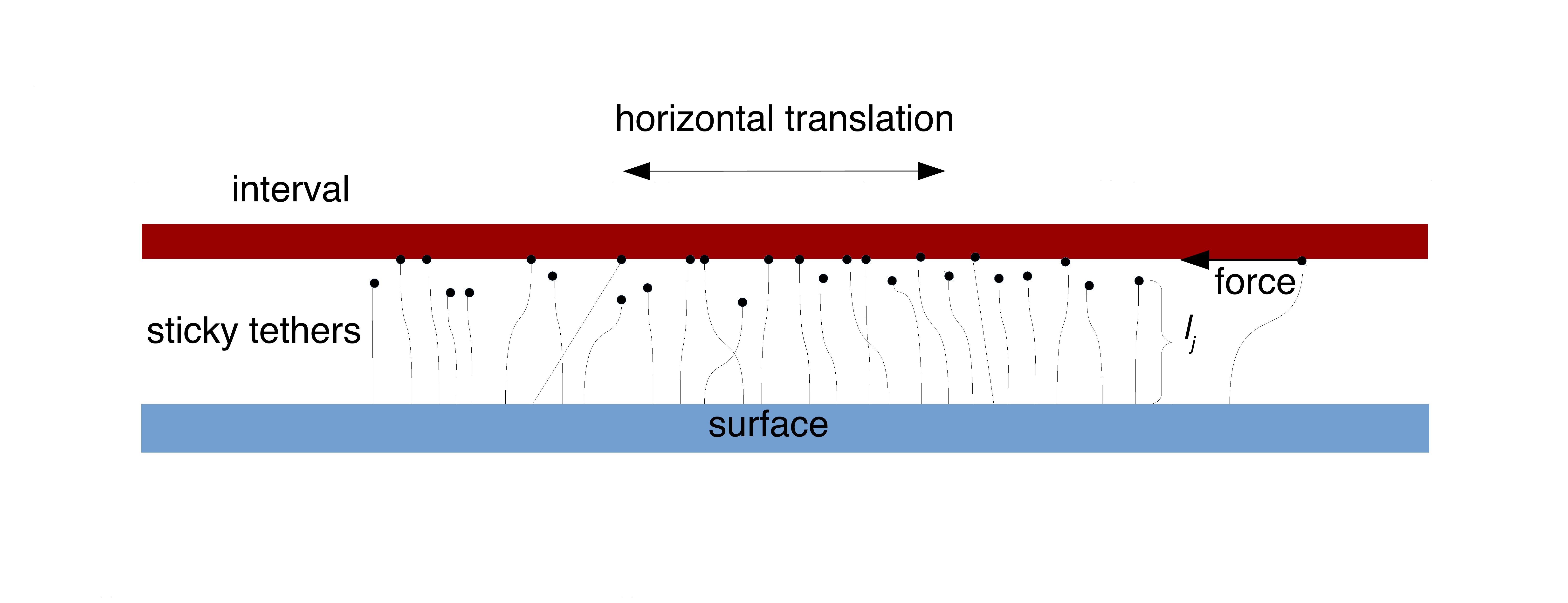}
\caption{
Schematic of the one-dimensional model of a particle moving on a DNA-coated surface. 
The particle is an interval that can translate horizontally in the $x$-direction. 
The surface is a line coated with sticky spring-like tethers of lengths $l_j$ that bind and unbind to the interval.  
The schematic is shown with extra vertical space for illustration but the interval and surface lie on the same horizontal line.
\label{1D_model}}
\end{figure}

We start by constructing a one-dimensional model for a particle moving near a DNA-coated surface. 
%For simplicity we consider only the surface to be coated with DNA. 
The model considers a one-dimensional interval of mass $m$ that can move horizontally along a one-dimensional line. 
Attached to the line are one or more sticky tethers that behave as thermal springs: they each have 
energy $\frac{1}{2} k(l-\bar l)^2$, where $l$ is their length, $\bar l$ is their mean rest length, and $k$ is the spring constant, and in the absence of any interaction with the interval, their lengths fluctuate stochastically at temperature $T$ according to overdamped dynamics. 
The tethers bind and unbind to the interval independently of each other with rates $q_{on}, q_{off}$. When the tethers are bound to the interval, they exert a horizontal force $-k(l-\bar l)$ on the interval. The interval's acceleration is given by the sum of the forces acting on it, as per Newton's law.  
The setup is sketched in Figure \ref{1D_model}.

In this one-dimensional model, the interval models the particle, such that the length of the interval is a proxy for the size of the patch on a spherical particle to which the surface's DNA can bind. 
The sticky tethers model the DNA which coats the surface, since a polymer in solution (such as DNA) behaves like a linear spring for small deviations from equilibrium \cite{Marko:wn}.  
The binding and unbinding models the sticky ends of DNA on different surfaces which hybridize. 

This model ignores a great many physical effects that could play a role. 
Nevertheless, it captures, with a minimum of parameters and assumptions, the two basic physical ingredients of the coarse-grained dynamics of the particle: friction and diffusion.   
To see how, suppose the interval is initially at rest with no external force. The tethers are constantly fluctuating and eventually a tether will bind to the interval and exert a force on it until the tether unbinds, causing the interval's velocity to change. As this and other tethers repeatedly bind and unbind at different lengths, each time exerting a force with a random magnitude and direction, the interval performs a random walk. Over long enough timescales this random walk approaches a Brownian motion with some DNA-induced diffusion coefficient $D_{\rm DNA}$.%\footnote{Note that technically, the velocity performs a random walk, but the velocity is also forced to the origin because of the additional friction force described in the next paragraph, so it is reasonable to think of the position as performing a random walk.}.

Now consider the dynamics of the interval when it is forced to translate horizontally at some velocity $v$, say to the right. Tethers still bind and unbind at random lengths, but when they are bound they are dragged to the right by the interval, so on average they exert a spring force to the left. For small perturbations the average force exerted by the tethers will be linear in the interval's velocity, as $-\Gamma_{\rm DNA}v$, where $\Gamma_{\rm DNA}$ is the DNA-induced friction coefficient. 

Our aim is to derive expressions for the DNA-induced friction and diffusion coefficients at long timescales, $\Gamma_{\rm DNA}$ and $D_{\rm DNA}$. 
If the system is in equilibrium then by the fluctuation-dissipation relation these coefficients are related  as $D_{\rm DNA}=k_BT\:\Gamma_{\rm DNA}^{-1}$, where $T$ is the temperature and $k_B$ is Boltzmann's constant. 
Our derivation doesn't assume the fluctuation-dissipation relation, although the fact that it holds after coarse-graining is a consistency check on the model.

 \subsection{Model setup}

The model is characterized by the following variables. 
The interval has velocity $v$, 
and there are $N\geq 1$ sticky tethers. 
Let $l_j(t)$ be the length of the $j$th tether at time $t$, and let $s_j(t)\in\{u,b\}$ be the state of the $j$th tether, where $s_j(t)=u$ if the tether is unbound at time $t$, and $s_j(t)=b$ if the tether is bound.
The complete set of variables is $\{v,\bfl =(l_j)_{j=1}^N, \bfs=(s_j)_{j=1}^N\}$. 

The energy of an unbound tether is $\half  k l^2$; we take the rest length to be $\bar l=0$ since it doesn't play a role in this one-dimensional model. We let the free energy difference between the unbound and the bound state of a tether be $-e_0$ (defined with a minus sign so that $e_0$ is positive when binding is energetically favourable.) The energy associated with the interval is purely kinetic. The energy of the entire system is 
\begin{equation}\label{E}
E = \half m v^2 + \sum_{j=1}^N \Big( \half kl_j^2  -e_0\delta_{s_j,b} \Big)\,.
\end{equation}
Here $\delta_{a,b}$ is a  Kronecker delta, equal to 1 if $a=b$ and 0 otherwise. 

The variables evolve in time according to the following dynamics, which we show in the ESI$^\dag$  preserve the Boltzmann distribution $\pi\propto e^{-\beta E}$, where $\beta=(k_B T)^{-1}$.

\begin{enumerate}
\item \emph{Tether binding/unbinding}

Tethers bind and unbind independently of each other with rate of binding $q_{on}$ and rate of unbinding $q_{off}$. That is, $\{s_j(t)\}_{j=1}^N$ is a collection of independent, continuous-time Markov chains with the given rates. 
 To satisfy detailed balance, the rates and binding energy must be related by 
\begin{equation}
 \label{binding_rates}
 q_{on} = q_{off} e^{\beta e_0} .
\end{equation}
In equilibrium, the  probabilities of being bound ($p_b$) or unbound ($p_u$) are$^\dag$
\begin{equation}\label{pbpu}
p_b = \frac{e^{\beta e_0} }{ (1 + e^{\beta e_0})}, \qquad p_u = \frac{ 1 }{ (1 + e^{\beta e_0})}\,.
\end{equation}

\item \emph{Tether length dynamics}

When the tethers are unbound, we model their dynamics with the overdamped Langevin equation, a reasonable model assuming small perturbations from equilibrium. 
We use a constant friction coefficient $\gamma$ to model the damping felt by the spring in an ambient fluid as the spring relaxes to its rest length. 

When the tethers are bound, they move at the velocity of the interval. The combined dynamics may be written formally as
\begin{equation}
\label{tether_SDE}
\dd{l_j}{t} = \left( -\frac{k}{\gamma}  l_j + \sqrt{\frac{2\beta^{-1}}{\gamma}} \eta_j(t) \right) \delta_{s_j,u} + (v\:  dt)\delta_{s_j,b} \,,
\end{equation}
for $j=1,\ldots,N$. 
Here $\{\eta_j(t)\}_{j=1}^N$ are $N$ independent white noises.

\item \emph{Interval dynamics}

The interval's acceleration equals the sum of the forces acting on it according to Newton's law:
\begin{equation}\label{interval_SDE}
m\dd{v}{t} = \sum_{j=1}^N (-kl )\:\delta_{s_j,b}\,.
\end{equation}

\end{enumerate}

The full model is given by \eqref{tether_SDE}, \eqref{interval_SDE}, plus the binding/unbinding dynamics.

Our model clearly contains many simplifications of the real system and we explain here some of our modeling choices. 

One is that we assume $q_{on}$ and $q_{off}$ are constant, and in particular that they are independent of the tether length. While it is reasonable to assume that $q_{on}$ is constant in a diffusion-limited system \cite{Rogers:2013dc}, $q_{off}$ is typically modelled as depending exponentially on length, via Bell's theory \cite{Bell:1978gn,Hanggi:1990en,Chen:2001kk}, implying an equal exponential dependence on length in the binding free energy $e_0$. %; one study even suggests the unbinding rate could be non-exponential \cite{Rogers:2013dc}. 
However, it has been shown that Bell's theory is too simple to describe the kinetics of DNA unbinding, and that in fact the binding/unbinding rates for DNA are roughly constant for small or moderate forces \cite{Ho:2009cr}. 
Therefore we neglect such length dependencies, assuming small enough forces or length perturbations that the variation in unbinding rates contributes negligibly to the formulas for $\Gamma_{\rm DNA}$, $D_{\rm DNA}$. Our neglect is also a mathematical convenience, since it allows us to obtain an analytic expression for $\Gamma_{\rm DNA}$, although in Section \ref{sec:1tether} we point out which parts of the analysis would change if the rates were length-dependent. 

Another is that we assume the tethers attach rigidly to the interval, so they are simply dragged along by the interval. A different model might attach the tethers to a spring on the interval, which exerts a force back on the tether. This would add additional degrees of freedom and complexity, that we felt would not change the final result to leading order, so have omitted this effect in this initial model.

We also do not account for the boundaries of the interval, which would require imposing a cutoff on the length at which a tether can bind, where the cutoff would depend on the relative position of the tether and the interval. Rather, we neglect space entirely, so the interval is actually an infinite line with some mass, and the tethers can be located anywhere, including in a single bundle. We do not expect the effect of boundaries to play an important role, and anyways they would be an artifact of the one-dimensional geometry.

We furthermore make no attempt to incorporate hydrodynamic interactions, either associated with the interval in a fluid, or between tethers and the interval. This is certainly an oversight, since the particles we wish to model are embedded in a fluid; indeed we have incorporated (albeit crudely) the hydrodynamic friction on each tether via the friction coefficient $\gamma$. It is tempting to include a similarly crude model for hydrodynamics by adding the Langevin-style terms $-\gamma_h v + \sqrt{2\beta^{-1}\gamma_h}\eta_h(t)$ to \eqref{interval_SDE}, where $\gamma_h$ represents some hydrodynamic friction on the interval. Technically there is no barrier to doing this and the analysis proceeds in the same way, leading to a final friction coefficient of $\Gamma_{\rm DNA}+\gamma_h$. 
However, it is well-known that because hydrodynamic velocity fluctuations decays slowly, often on the same timescale as velocity fluctuations of a particle in the fluid, there is no separation of timescales which allows the Langevin equation to correctly model the velocity correlations of a colloid in a fluid \cite{Hinch:1975ba,Roux:1992ic,Bian:2016hp}. 
A different option would be model the interval evolution directly in the overdamped limit, which is known to be valid over long enough timescales \cite{Roux:1992ic}. %However, we found it not straightforward to construct overdamped equations replacing \eqref{interval_SDE} that satisfied detailed balance, 
However, we do not wish to assume a priori that hydrodynamic fluctuations decay more quickly than the DNA-induced fluctuations. 

Therefore, we neglect hydrodynamics entirely, and focus instead on capturing the coarse-grained effect of the tethers in isolation. If it turns out, (as it will in some parameter regimes), that $\Gamma_{\rm DNA}$ calculated in this way is much larger than the friction induced by hydrodynamics, then this may still be a valid approximation. In parameter regimes without such a separation it will require more investigation to understand how to correctly weight $\Gamma_{\rm DNA}$ with the hydrodynamic friction.

\subsection{Nondimensional model and separation of scales ansatz}\label{sec:ansatz}

We now nondimensionalize the model and identify the small and large terms. 
There are 6 parameters:  
$k$ (kg/s$^2$), $\gamma$ (kg/s), $q_{on}$ (s$^{-1}$), $q_{off}$ (s$^{-1}$), $m$ (kg), $\beta$ (J$^{-1}$ = (kg$\cdot$m$^2$/s)$^{-1}$), 
three classes of variables $v$ (m/s), $l_j$ (m), $s_j$ (unitless), 
and three physical dimensions (mass, length, time). 

The natural scales in the model associated with the tethers are a typical lengthscale $r_{\rm DNA} \approx (k_BT/k)^{1/2}$, and a typical tether evolution time $t_{\rm DNA}$, measured either by the binding/unbinding timescales $q_{on}^{-1}$, $q_{off}^{-1}$, or by the correlation time in the overdamped dynamics, $\gamma/k$. We assume these two timescales have the same order of magnitude. 

We use the larger scales associated with the interval/particle to nondimensionalize the dynamics.
Suppose a typical particle has radius $R$; this doesn't enter anywhere in the equations but must be provided as a separate lengthscale of interest. 
Let $\tau = R^2/D$ be a typical time it takes a particle to diffuse its radius if its diffusion coefficient is $D$. We only determine a specific formula for $D$ after coarse-graining, but we may estimate it a priori as described momentarily. 
We nondimensionalize the dynamics using scales $m$ for masses, $R$ for lengths, and $\tau$ for times.

We may estimate $D$ in two different ways. If diffusion arises from a sum of random kicks of sizes $r_{\rm DNA}$ and durations $t_{\rm DNA}$, we would expect $D \approx r_{\rm DNA}^2 / t_{\rm DNA} \approx q_{off}k_BT/k$. Alternatively, we can first estimate the effective friction $\Gamma =\beta^{-1}D^{-1}$, which by dimensional analysis should be $\Gamma = F / L \cdot T$,  where $F$ is a typical force applied on the interval, $L$ is a typical length over which it is applied, and $T$ is a typical time over which it is applied. Substituting $F \approx Nk r_{\rm DNA}$, $L\approx r_{\rm DNA}$, $T\approx t_{\rm DNA}\approx q_{off}^{-1}$ gives an estimate $\Gamma\approx Nk/q_{off}$ so by the fluctuation-dissipation relation $D\approx q_{off}k_BT/Nk$.  
This differs from the first estimate by a factor of $N^{-1}$; later we will see that the second estimate is closer to the correct formula.

With these scales in hand we may proceed with our scaling ansatz. 
We consider a regime where the typical evolution time for a tether is much faster than the diffusion time for the particle, so that $t_{\rm DNA}/\tau \sim O(\eps^2)$, where $\eps \ll 1$ is a small dimensionless number. 
We furthermore assume that $r_{\rm DNA}/R \sim O(\eps)$, an ansatz that is consistent with the first and with diffusion arising because of kicks from the tethers, since $(r_{\rm DNA}/R)^2 = \eps^2(r^2_{\rm DNA}/t_{\rm DNA})/D$. 
These assumptions determine the scalings of $q_{on}^{-1}$, $q_{off}^{-1}$, $\gamma/k$, $l$, and $\beta\gamma l^2$. 

There are two additional timescales in the problem. 
One is $(v/l)^{-1}$, the evolution time of the bound tethers, determined by the magnitude of the velocity. 
We make the ansatz that this timescale is $O(\eps)$ compared to $\tau$, so that bound tethers exert their influence over long enough timescales to give rise to diffusion on longer timescales. Equivalently, we may assume the nondimensional velocity is order unity, $v/(R/\tau)\sim O(1)$, which makes sense if $v$ is the variable whose coarse-grained dynamics we seek. 

The final timescale is $(kl/mv)^{-1}$, which may be interpreted as the correlation time for velocity; again we assume this is $O(\eps)$ compared to $\tau$ so that we may observe long-time diffusion. 

Altogether the collection of scaling assumptions may be written as 
\begin{gather}
\frac{q_{on}^{-1}}{\tau}, \: \frac{q_{off}^{-1}}{\tau}, \frac{\gamma/\kappa}{\tau}\sim O(\eps^{2}), 
\quad \frac{\beta\gamma}{R^2/\tau} \sim O(1), 
\quad \frac{l}{R} \sim O(\eps),  \nonumber\\
  \frac{v}{R/\tau}\sim O(1), \quad \frac{(kl/m)^{-1}}{\tau^2/R} \sim O(\eps)\,. \label{scalings}
\end{gather}
We write $A\sim O(\eps^\alpha)$ for some dimensionless variable $A$ to mean that $A=\tilde A \eps^\alpha$, where $\tilde A\sim O(1)$ is some dimensionless number of order unity. %In what follows we won't explicitly write the tildes and nondimensional form, but rather will substitute scalings using \eqref{scalings}. 

The small parameter $\eps$ may be estimated from the approximate formula for $D$, as 
\begin{equation}\label{eps}
\eps^2 = \frac{t_{\rm DNA}}{\tau} \approx \frac{k_BT}{NR^2k} = \frac{1}{N}\left(\frac{r_{\rm DNA}}{R} \right)^2\,.
\end{equation}
Here $r_{\rm DNA}$ is a typical length fluctuation of a coiled DNA strand, not the end-to-end length of the DNA strand itself (which is an upper bound on the fluctuation.) For a typical microscale particle with a diameter of $R\approx$ 1$\mu$m coated with DNA tethers of end-to-end lengths $\approx50$nm, $\eps$ could be quite small.

\section{Effective friction and diffusion}\label{sec:effective}

When the timescale $\eps^2$ in \eqref{eps} is small, we may systematically average over the fast tether dynamics to obtain the dynamics of the interval's velocity over longer timescales.
We present and interpret the result in this section, saving the derivation for section \ref{sec:1tether} since it is not critical for understanding the result. 

The resulting coarse-grained equation for the interval's velocity $V(t)$ is the Langevin equation
\begin{equation}\label{SDEeffective}
m\dd{V}{t} = -\Gamma_{\rm DNA}V dt + \sqrt{2\beta^{-1}\Gamma_{\rm DNA}}\eta(t)\,,
\end{equation}
where the effective friction is
\begin{equation}\label{FrictionDNA}
\Gamma_{\rm DNA} = \frac{Nk}{1+e^{-\beta e_0}}\left( \frac{1}{q_{off}} + e^{\beta e_0} \frac{\gamma}{k}\right)\,.
\end{equation}
The effective diffusivity, equal to $\frac{1}{2}\lim_{t\to \infty}\frac{d}{dt}\langle X^2(t)\rangle$ where $X(t) = \int^t_0 V(s) ds$ is the interval's position, is 
\begin{equation}\label{DiffDNA}
D_{\rm DNA} = \beta^{-1}\Gamma^{-1}_{\rm DNA}\,.
\end{equation}

Equations \eqref{FrictionDNA}, \eqref{DiffDNA} are our main results. Given either of these, one can use them in a model to predict the evolution timescale and pathways of a system of DNA-coated colloids. While $\Gamma_{\rm DNA}$ may be directly compared to hydrodynamic friction, $D_{\rm DNA}$ may be easier to measure experimentally. For example, one could observe a DNA-coated colloid on a DNA-coated surface, and from the mean-squared displacement over long times infer $D_{\rm DNA}$, and then solve for $\Gamma_{\rm DNA}$ using \eqref{DiffDNA}.  

How should the expression \eqref{FrictionDNA} be interpreted? 
The terms in brackets have dimensions of time, each representing a natural timescale for the tether dynamics. The factor $q_{off}^{-1}$ is the average time for a bound tether to unbind. 
The factor $e^{\beta e_0}\gamma/k$ is the time for the tether to relax to equilibrium from a stretched position, counting only time during which it is bound -- in the absence of binding/unbinding dynamics, the timescale would be $\gamma/k$. However, because the tether can bind and unbind along the way, and it only relaxes when it is unbound, 
the time during which it is bound in the relaxation period is prolonged by a factor equal to the ratio of the average time bound to the average time unbound, %$p_b/p_u=$
$q_{off}^{-1}/q_{on}^{-1} = e^{\beta e_0}$. 
We can interpret the sum of these timescales, $\tau_b\equiv q_{off}^{-1}+e^{\beta e_0}\gamma/k$, as the ``bound'' correlation time, in other words the time it takes the tethers to reach equilibrium after a displacement, counting only time during which they are bound to the interval. 

The remaining factor, $Nk/(1+e^{-\beta e_0})$, has dimensions of force per length, and can be interpreted as the average change in force required to displace the system infinitesimally from equilibrium. 
Indeed, consider a system with one sticky tether in equilibrium, and an interval which is held still. Imagine freezing the tether in its current state (bound/unbound), displacing the interval by some distance $\Delta x$, and measuring the additional force required for this displacement (additional means the difference from the current force, since the net force will be zero in equilibrium.) If the tether was initially bound, it is stretched by an amount $\Delta x$ so the additional force is $F(\Delta x;b)=-k\Delta x$. If it was initially unbound, its length doesn't change so the additional force is $F(\Delta x;u)=0$. The average additional force over all lengths and states if the tether is in equilibrium is $-k\Delta x p_b$ where $p_b$ is the equilibrium probability the tether is bound (see \eqref{pbpu}).
With $N$ tethers, the average additional force would be $-Nk\Delta x p_b$. 
Dividing by $-\Delta x$ gives the desired factor.

We now show that this interpretation suggests a more heuristic, mean-field derivation of \eqref{FrictionDNA}. 
Suppose the interval moves at velocity $v$, and consider the average impulse it receives 
from the tethers over a time $\tau_b$, the bound correlation time. 
It is reasonable to assume that over this time period the states of the tethers are frozen, and at the end of the time period they instantly rearrange and form a new sample from the equilibrium distribution. 
A single tether whose state is constant over the time period will supply an additional impulse of about $F(\Delta x;l,s)\cdot \tau_b$, where $\Delta x\approx v/\tau_b$ is the interval's displacement over the time period, and 
$F(x;l,s)$ is the additional force on the interval when the states of the tethers are frozen and the interval is displaced some amount $x$. 
The whole collection of tethers will therefore supply an additional impulse of $N \langle F(\Delta x;\bfl,\bfs)\rangle_{\pi_0} \cdot \tau_b$, where $\pi_0\equiv Z^{-1}_{\bfl,\bfs}\int_v\pi(\bfl,\bfs,v)dv$ is the marginal stationary distribution for the tethers in $\bfl,\bfs$ (or the distribution for fixed $v$; they are the same in this model) with $Z^{-1}_{\bfl,\bfs}$ the normalizing constant. 
For a system near equilibrium the average frictional force will be linear in velocity and with the opposite sign, so the average impulse has the form $-\Gamma_{\rm DNA} v\:\cdot\: \tau_b \approx -\Gamma_{\rm DNA}\Delta x$, for some positive coefficient $\Gamma_{\rm DNA}$. 
Equating these two formulas for the average impulse gives the following heuristic equation for calculating the DNA-induced friction: 
\begin{equation}\label{heuristic}
\Gamma_{\rm DNA} \approx N\left\langle \frac{\partial F}{\partial x} \right\rangle_{\pi_0} \cdot\; \tau_b\,.
\end{equation}

Formula \eqref{heuristic} is exact for the one-dimensional model when $\tau_b$ is defined as above. 
We expect it to also apply in higher dimensions, where $F$ and $x$ will be vectors, 
and $\Gamma_{\rm DNA}$ will be a matrix such that the frictional force vector when the system moves with velocity vector $v$ is $-\Gamma_{\rm DNA}v$. 
The higher-dimensional version can be justified using a similar heuristic argument as above, by considering the impulse in each of a set of orthogonal directions, to calculate the columns of $\Gamma_{\rm DNA}$ via \eqref{heuristic}. Of course, the argument does not tell us what the timescale $\tau_b$ should be, but if the timescale is roughly the same for all directions, then we can compare the relative magnitudes of the friction coefficients along different directions. We will do this for a specific higher-dimensional model in section \ref{sec:2d}.

Notice the similarity of \eqref{FrictionDNA} to the formula $\Gamma \approx Nk/q_{off}$ guessed in section \ref{sec:ansatz} by dimensional analysis. The correct formula has a smaller constant $N/(1+e^{-\beta e_0})$, which could have guessed, an an additional factor $e^{\beta e_0} \gamma / k$, related to the relaxation dynamics of the tethers, whose exact form would have been harder to guess. Indeed the formula from dimensional analysis has been derived in other ways in the study of sticky processes in biology, like dynein binding and unbinding to a microtubule giving rise to protein friction \cite{Tawada:1991uz}. However, the formula from dimensional analysis gives results that do not always agree  
 with our numerical simulations (Section \ref{sec:numerics}), which is why we needed to perform formal homogenization to obtain the correct formula.

\subsection{Numerical simulations}\label{sec:numerics}

\begin{figure}
 \centering
 \includegraphics[width=0.9\columnwidth]{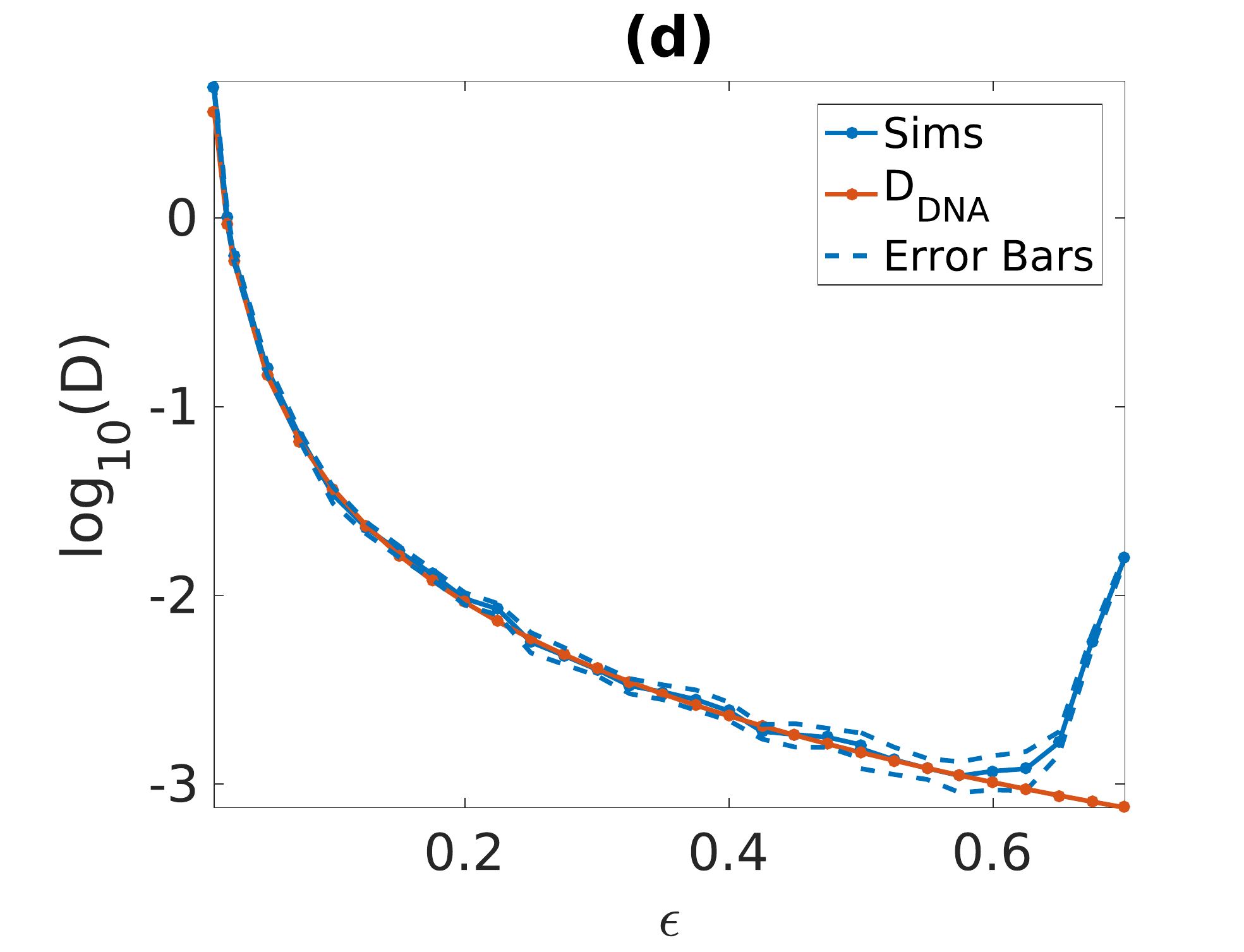}
 \caption{
 Simulations of the one-dimensional dynamics for non-infinitesimal parameter values. 
 %(a-d) Mean-squared displacement versus time for different $\eps$. 
%The corresponding diffusion coefficients are:  (a) $\eps=0.1$,  $D_{\rm numerical} = 0.039 \pm 0.003\ m^2/s$, $D_{\rm DNA} = 0.037\ m^2/s$; 
 %(b) $\eps=0.05$, $D_{\rm numerical} = 0.15 \pm 0.01\ m^2/s$, $D_{\rm DNA} = 0.15\ m^2/s$;
%(c) $\eps=0.02$, $D_{\rm numerical} = 0.97 \pm 0.07\ m^2/s$, $D_{\rm DNA} = 0.92\ m^2/s$.
 %(d)  
Logarithm of diffusivity from numerical simulations $\log(D_{\rm numerical})$ (solid blue) and theoretical prediction $\log(D_{\rm DNA})$ (orange) as a function of $\eps$. Error bars for the numerical estimate (dashed blue) are one standard deviation from the mean.
\label{diffusivity_eps}}
\end{figure}

We verified that our asymptotic analysis is accurate for a system with non-infinitesimal parameters by numerically simulating the model and comparing the interval's velocity statistics with those predicted by the coarse-grained equation. 
We simulate a system of $N=1000$ tethers using the following parameters:
\begin{gather}
m = \eps^2, \; q_{off} = \eps^{-2}, \; \beta =e_0 = 1, %\nonumber\\
q_{on} = e\eps^{-2}, \; \gamma = \eps^2, \; k = 1\,.
\end{gather}
Here $\eps$ is a numerical parameter that we vary. One can verify that when $\eps \ll 1$, the generator has the scaling required for the asymptotic analysis to work, (see \eqref{eps_generator}), provided  that $v\sim O(\eps^{-1})$. 
We tried other parameter combinations that scaled the same way with $\eps$ and there was little difference in the quality of the comparisons. 

Equations \eqref{tether_SDE}, \eqref{interval_SDE} are solved using the Euler-Maruyama method with timestep $\Delta t = 0.002\epsilon$, and the binding/unbinding dynamics are incorporated by allowing each tether to change state during each timestep with probability $q\Delta t$, where $q\in\{q_{on},q_{off}\}$ depends on which state the tether is currently in. 
The position $x(t)$ of the interval is obtained by integrating the velocity using the same first-order method.

To collect statistics we repeated each simulation 100 times to obtain an ensemble of realizations, and computed 10 ensembles for each parameter value to calculate error bars. 
The diffusion constant is computed for each ensemble as the slope of the least squares-fit line to the mean-squared interval displacement, $\inner{x^2(t)}$, over the period $t\in [10,20]$; early time points are thrown away to avoid fitting to the transient behaviour. 
The reported effective diffusion constant $D_{\rm numerical}$ is the mean over the 10 ensembles.

Figure \ref{diffusivity_eps} shows that  the numerically computed diffusivity agrees with the formula \eqref{DiffDNA} for $\eps\lessapprox0.6$. For $\eps \gtrapprox 0.6$, the simulations and theory diverge, with the simulation diffusivity becoming significantly larger. 

%Figure \ref{diffusivity_eps} (a) plots the mean-squared displacement $\inner{x^2(t)}$ for three  values of $\eps$, verifying that the behaviour is diffusive. Figure  \ref{diffusivity_eps} (b) compares the numerically computed diffusivity with that predicted by formula \eqref{DiffDNA} for a range of $\eps$; the two agree within error bars even for values as large as $\eps=0.5$. For $\eps \gtrapprox 0.6$, the simulations and formula \eqref{DiffDNA} differ, with the simulation diffusivity becoming significantly larger. 

\section{Estimating the physical magnitude of DNA-induced friction}\label{sec:physics}

 \begin{figure}
  \centering
  \includegraphics[width=\columnwidth]{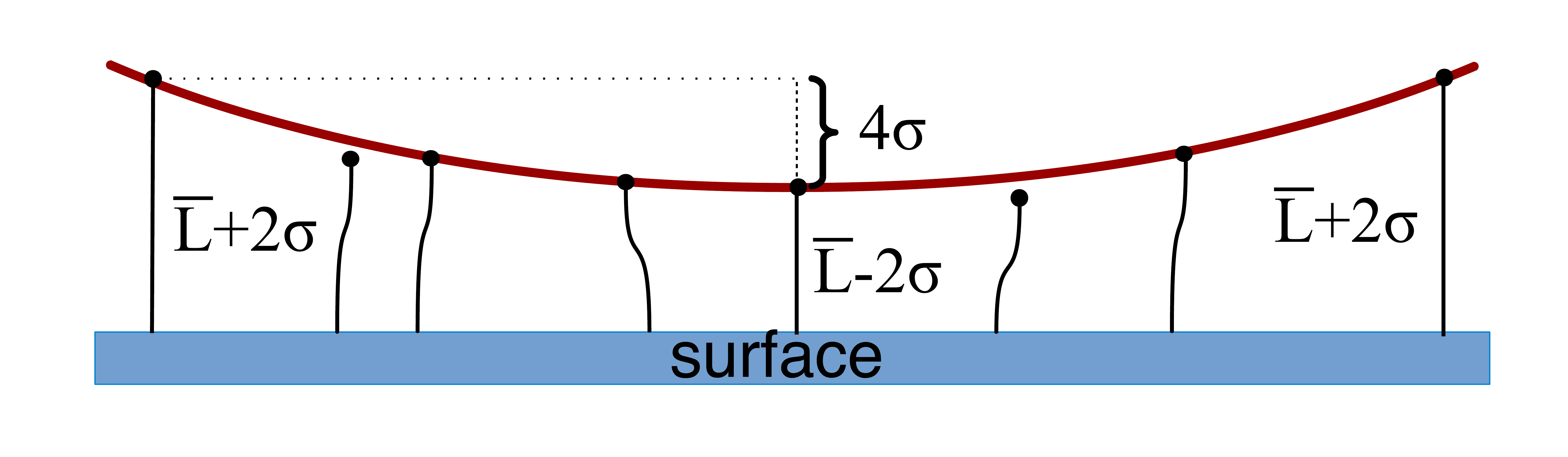}
 \caption{ 
 The patch of DNA strands on a spherical DNA-coated colloid which can interact with strands on a surface forms a spherical cap with height $h$, roughly equal to 4 times the standard deviation of the length of a tether.  $\bar L$ is the rest length of the tether. 
 \label{particle_fraction}}
 \end{figure}

We use \eqref{FrictionDNA} to estimate the magnitude of DNA-induced friction for a  microscale colloid moving on a surface coated densely with weakly-interacting DNA strands, and compare it to the hydrodynamic friction that it must also experience. 
%felt by a colloid as it moves in a fluid. If DNA-induced friction is much smaller than hydrodynamic friction, then we may safely neglect it and focus on understanding hydrodynamic interactions instead. If it is much larger than hydrodynamic friction, then we may be justified in developing more detailed models of DNA-induced friction, and perhaps even neglecting hydrodynamics altogether. 
%
%Let us estimate $\Gamma_{\rm DNA}$ for a typical system that allows microscale colloids to move relative to each other. 
We consider a particle with a radius of $R=0.5\mu$m at a temperature of about $T=45^\circ$C (318$^\circ$K), and with single-stranded DNA (ssDNA) of length $L= 40$nm (about 65 base pairs \cite{Rogers:2011et}), similar to the system considered in Wang et al\cite{Wang:2015ep}, in which  DNA-coated colloids crystallized. 
We base our estimates for physical properties of the tethers on the worm-like chain (WLC) model for a flexible polymer \cite{Marko:wn}. 
For small perturbations from equilibrium the linear spring constant in the WLC model is
\begin{equation}\label{kest}
k = \frac{3k_BT}{2AL} \approx 6.6\times 10^{-5}\:\text{kg/s} \approx 10^{-5}\:\text{kg/s}\,,
\end{equation}
where we have substituted the persistence length for ssDNA to be about $A=2.5$nm \cite{Murphy:2004gn}. Recall that $k_B=1.38\times 10^{-23}$m$^2$kg$\cdot$s$^{-2}$K$^{-1}$. 

One caveat here is that the WLC model considers a polymer in free space, while our polymer is tied down at one end and constrained to live between two surfaces in close proximity, which could change its thermodynamics. 

For the friction coefficient $\gamma$ we assume the drag a polymer feels when its length is changing at a given velocity, is approximately equal to the drag it would feel if it were moving through a fluid with fixed length at that velocity. 
For small length perturbations the polymer is approximately spherical so we use the Stokes' drag on a sphere, $6\pi\eta R_H$, where $\eta$ is the fluid viscosity, equal to $8.9\times 10^{-4}$kg/(m$\cdot$s) for water, and $R_H$ is the hydrodynamic radius of a polymer, equal to $R_H=0.375(AL)^{1/2}$ in the WLC model \cite{Marko:wn}. The result is 
\begin{multline}\label{gammaest}
\gamma \approx 6\pi\eta \cdot 0.375(AL)^{1/2} \approx 6.3\times 10^{-11} \text{kg/s}^2\approx 10\times 10^{-10} \text{kg/s}^2 \,.
\end{multline}
This may be an overestimate, since the tether will not be perfectly spherical and drag on an ellipse in the long direction is strictly less than that on a sphere \cite{Marko:wn}, but this drag term won't contribute much. 

For the binding and unbinding rates, $q_{on},q_{off}$, we refer to Bonnet et al\cite{Bonnet:1998gy}, which measured binding/unbinding rates for a DNA hairpin loop in solution, similar to ssDNA binding/unbinding. They found the binding rate $q_{on}$ was relatively insensitive to temperature \cite{Bonnet:1998gy} (e.g. Figs 4--6 in their paper), in the range of $10^4-10^5$s$^{-1}$ for different loop lengths and salt concentrations, while the unbinding rate $q_{off}$ depended strongly on temperature, varying from $10^2$s$^{-1}$ at low temperatures to $10^5$s$^{-1}$ at high temperatures, though it was relatively insensitive to loop length and salt concentration. 
To be conservative and underestimate the friction, we use the upper bounds for $q_{on}$ and $q_{off}$ and a ratio $q_{on}/q_{off}$ that is slightly smaller than the real one (due to a higher $q_{off}$); this underestimate partly compensates for not incorporating any length-dependence in the unbinding rate. Our rates and binding energy are
\begin{equation}\label{ratesest}
q_{on}\approx 10^5\:\text{s}^{-1},\quad q_{off} \approx 10^5\:\text{s}^{-1}\,,\quad e^{\beta e_0} \approx 1\,.
\end{equation}
These choices of rates imply a small $|\beta e_0|$, which is reasonable for weakly interacting DNA strands.

The remaining parameter to estimate is $N$, the number of DNA strands on the sphere that can reach and bind to DNA strands on the surface. This interaction patch of accessible DNA strands forms a spherical cap of height $h$ to be determined (Figure \ref{particle_fraction}.) 
Assuming that DNA strands may fluctuate in length by about $\pm$ 2 standard deviations to bind to neighbours, the maximum height difference between the bottom of the spherical cap and the top is 4 standard deviations, or $h = 4/\sqrt{\beta k}$. 
Multiplying the area of a spherical cap, $2\pi Rh$, by the DNA density, which we take to be $\rho = $ 1 DNA per $27$nm$^2$ as in Wang et al\cite{Wang:2015ep}, and substituting for $k$ from \eqref{kest}, gives
\begin{equation}\label{Nest}
N \approx 2\pi R\cdot \frac{4}{\sqrt{\beta k}}\cdot \rho  = 3 \times 10^3\,.
\end{equation}
Since there are $10^5$ DNA strands total on a particle of radius $0.5\mu$m, this estimate suggests about 3\% of the DNA strands are in the interaction patch at any given time. %, consistent with intuition; without any calculations we might estimate the patch to contain about 1-5\% of the strands. 

Plugging in the estimates \eqref{kest},\eqref{gammaest},\eqref{ratesest},\eqref{Nest} into \eqref{FrictionDNA} gives
\begin{align}
\Gamma_{\rm DNA} &\approx \frac{3\times 10^3}{1+1}\times 10^{-4}\frac{\text{J}}{\text{m}^2} \times \left( 10^{-5}\text{s} +\frac{10^{-10}}{10^{-4}}\text{s}\right) \nonumber \\
&\approx 2\times 10^{-6}\:\frac{\text{N} \cdot \text{s}}{\text{m}}\,.
\end{align}

Now we compare this to the hydrodynamic friction felt by a sphere with the same radius. In free space, the friction coefficient is $6\pi\eta R = 8.3\times 10^{-9} $kg/s.  Near a wall, the friction is about two times bigger \cite{Perry:2015ku}, so the friction coefficient is about $\gamma_{hydro} \approx 2\times 10^{-8}$kg/s. 
The ratio is
\begin{equation}
\frac{\Gamma_{\rm DNA}}{\gamma_{hydro}} \;\;\approx\;\; \frac{2\times 10^{-6}\:\text{kg/s}}{2\times 10^{-8}\:\text{kg/s}} \;\;\approx\;\; 100 \,.
\end{equation}
With these parameter values we estimate DNA-induced friction to be about 100 times bigger than the background hydrodynamic friction. Therefore, we \emph{do} expect DNA to have a significant effect on the kinetics of DNA-coated colloids under certain conditions, and perhaps an \emph{even greater effect than hydrodynamics}. 
We also expect the kinetics of DNA-coated colloids to be significantly slower than those of colloids interacting via other mechanisms such as depletion, for example, since the diffusion coefficient may be about 100 times smaller. 

%It is interesting to consider the friction as a function of the natural parameters of the system, the radius of the particle $R$ and the length of a DNA strand $L$. Substituting our models for each of the parameters gives
%\begin{align}
%\Gamma_{\rm DNA} %&= \frac{8\pi(3/2)^{1/2}}{A^{1/2}}\frac{1}{\beta(1+e^{-\beta e_0})} \;\frac{R\rho}{L^{1/2}} %\nonumber \\
%%&\quad\times
% %\left ( q_{off}^{-1} + \frac{9\pi\eta A^{1/2}}{4} \:e^{\beta e_0}L^{1/2} \right) \nonumber \\
%&= C_1\; \frac{R\rho}{L^{1/2}} \left ( q_{off}^{-1} + C_2 e^{\beta e_0} \:L^{1/2} \right)\,.
%\end{align}
%Here $C_1,C_2$ are constants; $C_1$ is temperature-dependent. 
%The friction decreases as the length of the DNA increases, unless the binding energy or fluid viscosity are large enough that the DNA relaxation timescale dominates, in which case the friction is independent of length. The friction increases linearly with the radius of the particle if the DNA density is constant. If $N$, the number of tethers in the interaction patch, is held constant, for example to keep the strength of the effective interaction constant, then the friction decreases as $L^{-1}$ and is independent of particle radius. Therefore, neglecting the length dependence of binding/unbinding kinetics, this model suggests that to increase the mobility of DNA-coated colloids one should use longer DNA strands. 

\section{Toward a two-dimensional model}\label{sec:2d}

\begin{figure}
\begin{center}
\begin{tikzpicture}[scale=1]

% parameters
\pgfmathsetmacro{\R}{2};          % radius of circle
\pgfmathsetmacro{\h}{0.75*\R}  % height of disc
\pgfmathsetmacro{\th}{23}         % angle of tether
\pgfmathsetmacro{\y}{0.3*\R}     % y-location of tether
\pgfmathsetmacro{\rarc}{0.45*\R}  % radius of angle arcs
\pgfmathsetmacro{\rphiarc}{1.3*\R}  % radius of phi arc
\pgfmathsetmacro{\xiang}{48}        % \xi -- must guess for each parameter combo by eye

% draw basic elements
\coordinate (A) at (0,\R+\h);
\coordinate (C) at (0,\h);
\draw[name path=disc] (A) circle [radius=\R];            % disc
\draw (-\R*1.5, 0) -- (\R*1.5,0);       % line
\draw[dashed] (A) -- (C);               % dashed line to contact point
\filldraw[black] (A) circle (2pt);                 % center of circle
\filldraw[blue] (C) circle (2pt) node[align=center,anchor=south east] {C};    % contact point C 
\draw (0,3pt) -- (0,-3pt) node[align=center,below=2pt] {$x$};     % x-tick
\filldraw[black] (\y,0) circle (1.5pt);                 % y-point
\draw (\y,3pt) -- (\y,-3pt) node[align=center,below=2pt] {$y$};     % y-tick
\draw[brace] (-0.05,0.05) --node[left=3pt] {$h$} (-0.05,\h-0.05);   % brace for height

% calculate points B,D and draw lines involving them
\path [name path=tether] (\y,0)--({\y+1.6*\R*sin(\th)},{1.6*\R*cos(\th)});   % "virtual" tether path
\path [name path=vertical] (\y,0)--(\y,\R);                                            % "virtual" vertical tether
\draw[name intersections={of=tether and disc, by=B},wiggly,violet] 
        (\y,0) -- (B) node[pos=0.5,right=3pt] {$l$} ;   % draw tether 
\filldraw[violet] (B) circle (2pt);                 % tether binding point
\draw[dashed, name intersections={of=vertical and disc, by=D},violet] (\y,0) -- (D);   % vertical tether
\draw[dashed] (A)-- (B) node[above=6pt,pos=0.66] {$R$}  ;    % line from disc center to binding point 
\node[violet,anchor=north west] at (B) {$B$};

% draw angle arcs
%            see: http://www.texample.net/tikz/examples/three-link-annotated/
% \theta
\draw[->,violet] ({\y+\rarc*sin(\th)},{\rarc*cos(\th)})  arc[radius=\rarc,start angle=90-\th, end angle=90] ;
\node[violet,above,align=center] at ({\y+\rarc*sin(\th/2)},{\rarc*cos(\th/2)}) {$\theta$};
% \xi
\draw[->] (0, \R+\h-\rarc)  arc[radius=\rarc,start angle=270, end angle=270+\xiang] ;
\node[below,align=center] at ({0+\rarc*sin(\xiang/2)},{\R+\h-\rarc*cos(\xiang/2)}) {$\xi$};

% draw phi arc
\draw[->] (\rphiarc, \R+\h) arc[radius = \rphiarc*0.7,start angle = -10, end angle = 35];
\node[right=3pt] at (\rphiarc, \R+\h + \R*0.4) {$\phi$};

\end{tikzpicture}
\end{center}
\caption{Schematic of a disc moving above a DNA-coated line. A disc of radius $R$ sits at fixed height $h$ (distance from line to the ``contact point'' $C$) above a line, at variable horizontal position $x$ and overall rotation $\phi$. A tether of variable length $l$ (violet) has one end on the line at fixed position $y$, and the other end bound at ``binding point'' $B$ on the disc at variable angle $\theta$ to the vertical, forming an angle $\xi$ with the center of the disc.}
\label{fig:disc}
\end{figure}
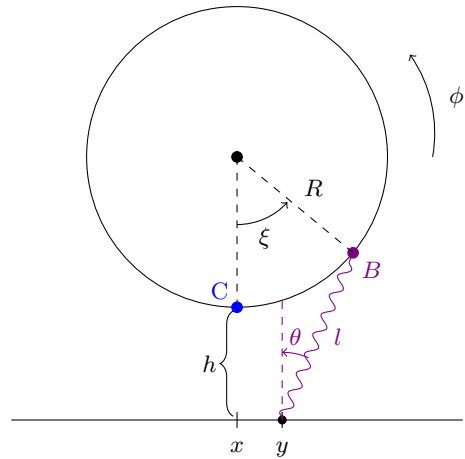

We now consider how the one-dimensional model may be extended to higher dimensions. 
While fully setting up and coarse-graining a higher-dimensional model is beyond the scope of this paper, we may still appeal to the heuristic equation for the friction tensor $\Gamma_{DNA}$ \eqref{heuristic}, 
which we justified in a mean-field manner by considering the average impulse applied by the tethers over a typical tether correlation time $\tau_b$. We argued that one can use \eqref{heuristic} to estimate the components of the friction tensor up to an unknown timescale $\tau_b$, and if this timescale is roughly the same for all components of the friction tensor, then we may still compare the relative magnitude of the friction coefficients for different directions of motion. We are particularly interested in determining whether some directions have significantly more friction than others, like sliding compared to rolling, as suggested by the velcro analogy in the introduction. 

Let us consider a two-dimensional disc of radius $R$ moving near a line coated with sticky tethers (Figure \ref{fig:disc}.) The disc can translate horizontally and rotate; we ignore motion in the vertical direction. The tethers are modeled as springs with lengths $l$, angles from the vertical $\theta$, and energies 
\begin{equation}\label{Etether2d}
\Et = \frac{1}{2}k_l(l-\bar l)^2 + \frac{1}{2} k_{\theta}\theta^2\,.
\end{equation}
Here $k_l$ is the spring constant associated with stretching the tether, and we have included a bending energy with spring constant $k_\theta$ to penalize deviations from vertical. This additional term is a crude attempt to model the stiffness of the anchors binding DNA strands to surfaces, as well as steric interactions which encourage the closely-packed DNA strands to stand up straight. 

Consider a disc whose point closest to the line, the ``contact'' point C, has height $h$ above the line and horizontal position $x$, and whose overall rotation is measured by angle $\phi$. Let there be a tether with length $l$, angle $\theta$, and horizontal position $y$. Suppose the tether is bound to the disc at point $B$, the ``binding'' point, which has angle $\xi$ from the vertical relative to the disc's center, as in Figure \ref{fig:disc}. The length and angle variables are related by considering the horizontal and vertical coordinates of the attachment point:
\begin{equation}
y-x+l\sin\theta = R\sin \xi\,, \qquad
R(1-\cos \xi) + h = l\cos \theta \,. \label{attach}
\end{equation}
Given three variables from the set $\{l,\theta,\xi,x,h\}$ one can use \eqref{attach} to solve for the other two. Therefore, the energy of a bound tether can also be considered a function of $(x,\xi)$ with $h$ entering as a parameter.

To use \eqref{heuristic} we must determine the additional force associated with translating and rotating the disc, i.e. moving it in the directions corresponding to $x,\xi$, for each possible binding point $B$. 
This additional force is found from the tether energy as $\mathbf{F} = -\nabla_{x,\xi} \Et$; we substituted $\phi\to\xi$ since $\partial_\phi = \partial_\xi$.  Note the second component of the force vector is a torque, with units of force $\cdot$ distance. The friction tensor predicted heuristically from \eqref{heuristic} would therefore be 
$\Gamma_{\rm DNA} = \langle \Lambda \rangle_{\pi(l,\theta,s;y)\rho(y)}\cdot \tau_b$, with the infinitesimal forces when $s=b$ equal to 
\begin{equation}\label{Lambda}
\Lambda = 
\begin{pmatrix}  
\frac{\partial^2\Et}{\partial x^2} & \frac{\partial^2\Et}{\partial x\partial \xi} \\
\frac{\partial^2\Et}{\partial x\partial \xi} & \frac{\partial^2\Et}{\partial \xi^2} 
\end{pmatrix}\,.
\end{equation}
Here $\rho(y)$ is the number density of DNA strands on the line, and $\pi(l,\theta,s;y)$ is the stationary distribution for the tether when the disc is fixed in place. This distribution will be related to $e^{-\beta \Et}$ but not exactly equal, because for $s=b$ only one of $l,\theta$ is independent by the constraints \eqref{attach}. 

We may compute each of the components of $ \Lambda$ for a tether fixed at $y$ and bound at angle $\theta$ by differentiating $\Et$ with respect to $x,\xi$ using the chain rule, and substituting for derivatives $\partial (l,\theta)/ \partial (x,\xi)$, $\partial^2(l,\theta) / \partial(x,\xi)^2$. We computed these derivatives by  differentiating \eqref{attach} implicitly using Mathematica; the resulting expressions are shown in the ESI$^\dag$. 
After redefining $\xi \to R\xi$ so the components of $ \Lambda$ have the same units, 
we obtain the infinitesimal forces %, shown in Table \ref{tbl:2dFriction}. 
%\begin{table*}
%%\twocolumn[ \begin{@twocolumnfalse}
%\begin{multline}\label{2dFriction}
%\Lambda = k_l \begin{pmatrix}
%\frac{\Delta l}{l} + (1-\frac{\Delta l}{l})\sin^2\theta & 
%\sin\theta\sin(\xi+\theta) + \frac{\Delta l}{l}\cos\theta\cos(\xi+\theta) \\
%\sin\theta\sin(\xi+\theta) + \frac{\Delta l}{l}\cos\theta\cos(\xi+\theta) & 
%\;\;\sin^2(\xi+\theta) + \frac{\Delta l}{l}\cos(\xi+\theta)\big(\frac{l}{R}+\cos(\xi+\theta) \big)
%\end{pmatrix}\\
%+ \frac{k_\theta}{l^2} \begin{pmatrix}
%\cos^2\theta + \theta \sin 2\theta & 
% \cos\theta\cos(\xi+\theta) - \theta\sin(\xi+2\theta)\\
% \cos\theta\cos(\xi+\theta) - \theta\sin(\xi+2\theta) & 
%\;\; \cos^2(\xi+\theta) + \theta\sin 2(\xi+\theta) + \frac{l}{R}\theta\sin(\xi+\theta)
%\end{pmatrix}
%\,.
%\end{multline}
%% \end{@twocolumnfalse} ]
%\caption{Full expression for the infinitesimal forces associated with a tether fixed at $y$ on the surface and bound at angle $\theta$. The other variables, $\xi$ and $\Delta l = l-\bar l$, are determined by the constraints \eqref{attach}.}\label{tbl:2dFriction}
%  \end{table*}
%Here $\Delta l = l-\bar l$ is the deviation from a tether's rest length. 
%Although $\Lambda$ is written as a function of $\theta,l,\xi$, only one of these variables may be chosen independently for a given tethering point $y$, by the constraints \eqref{attach}. 
%
%
\begin{widetext}
\begin{multline}\label{2dFriction}
\Lambda = k_l \begin{pmatrix}
\frac{\Delta l}{l} + (1-\frac{\Delta l}{l})\sin^2\theta & 
\sin\theta\sin(\xi+\theta) + \frac{\Delta l}{l}\cos\theta\cos(\xi+\theta) \\
\sin\theta\sin(\xi+\theta) + \frac{\Delta l}{l}\cos\theta\cos(\xi+\theta) & 
\;\;\sin^2(\xi+\theta) + \frac{\Delta l}{l}\cos(\xi+\theta)\big(\frac{l}{R}+\cos(\xi+\theta) \big)
\end{pmatrix}\\
+ \frac{k_\theta}{l^2} \begin{pmatrix}
\cos^2\theta + \theta \sin 2\theta & 
 \cos\theta\cos(\xi+\theta) - \theta\sin(\xi+2\theta)\\
 \cos\theta\cos(\xi+\theta) - \theta\sin(\xi+2\theta) & 
\;\; \cos^2(\xi+\theta) + \theta\sin 2(\xi+\theta) + \frac{l}{R}\theta\sin(\xi+\theta)
\end{pmatrix}
\,.
\end{multline}
\end{widetext}
Here $\Delta l = l-\bar l$ is the deviation from a tether's rest length. 
Although $\Lambda$ is written as a function of $\theta,l,\xi$, only one of these variables may be chosen independently for a given tethering point $y$, by the constraints \eqref{attach}. 

Given a model for the binding energy, one can average $\Lambda$ in \eqref{2dFriction} with respect to the stationary distribution $\pi(l,\theta,s;y)$ and then integrate over the number density $\rho(y)$ to obtain the effective friction tensor. We will not do this here, because constructing and verifying a fully consistent two-dimensional energy model is beyond the scope of the paper. However, we can use \eqref{2dFriction} to estimate the leading-order contribution to the friction tensor when $\bar l/R\ll 1$ and the DNA is very stiff. In this case most of the bound tethers will be near the contact point C and will bind nearly vertically close to their rest lengths so $\xi,\theta,\Delta l/\bar l, (y-x)/\bar l$ will all be very small. 
Specifically, if $\theta \ll 1$ then we find from \eqref{attach} that 
$(y-x)/\bar l \sim O(\theta)$, $\Delta l/\bar l \sim O(\theta^2)$, $(y-x)/\bar l\sim O(\theta)$, $\xi \sim O(\theta \bar l/R)\ll \theta$. 

We expand \eqref{2dFriction} in powers of these small parameters to obtain the infinitesimal forces up to $O(\theta^2)$ as
\[
 \Lambda_2  = 
k_l\Big(\frac{\Delta l}{\bar l} + \theta^2\Big)\begin{pmatrix}
1 & \;\;1 \\ 
1 & \;\;1
\end{pmatrix}
\;+\;
\frac{k_\theta}{\bar l^2}\begin{pmatrix}
1+\frac{3}{2}\theta^2 & \;\;1 -3\theta^2\\ 
1-3\theta^2 & \;\; 1+\frac{3}{2}\theta^2
\end{pmatrix}\,.
\]
The friction force to $O(\theta^2)$ for some velocity $v=(x'(t),R\phi'(t))$ is obtained from this matrix as $(f_1,f_2) = -\Lambda_2 v\cdot \tau_b$. The first component $f_1$ is a force in the $x$-direction, and the second component $f_2$ is the torque divided by $R$. 
Notice that to leading order, i.e. ignoring terms of $O(\theta^2)$, the matrix $\Lambda_2$, is degenerate, with null space spanned by $u_r=(1,-1)^T$. Since in this direction $(f_1,f_2) = -\Lambda_2 u_r\cdot \tau_b= (0,0)$, there is no friction to leading order for motion in this direction. 

What kind of motion is the low-friction direction $u_r$? This corresponds exactly to the disc \emph{rolling} along the line, since the disc rolls when the velocity of the contact point C, $x'(t) + R\phi'(t)$, equals zero. 
The disc is like a gear, and the line is like a gear track, and the disc rolls commensurately along the line without slipping. Therefore, we expect significantly smaller friction for rolling than for other kinds of motions, approaching zero as the DNA becomes shorter and stiffer. 

To say how much smaller the friction for rolling should be we compute the eigendecomposition of $\Lambda_2$, since the eigenvector with the smallest eigenvalue is the direction with the smallest friction (and the eigenvalue is the friction coefficient for that direction), and the eigenvector with the largest eigenvalue is the direction with the largest friction. 
The eigenvalues are 
\begin{equation*}
\lambda_r=\frac{k_\theta}{\bar l^2}\cdot \frac{9}{2}\theta^2,\;\quad
\lambda_s=2k_l\Big(\frac{\Delta l}{\bar l} + \theta^2\Big) + \frac{k_{\theta}}{\bar l^2}\Big(2-\frac{3}{2}\theta^2\Big)
\end{equation*}
and the respective eigenvectors are
\begin{equation*}
u_r=(1,-1), \;\quad u_s=(1,1)\,.
\end{equation*}
The rolling direction $u_r$ remains the eigenvector with the smallest eigenvalue even to $O(\theta^2)$.
Even then, the eigenvalue $\lambda_r$ only depends on the bending energy; the stretching energy still does not contribute at this order.  

The direction with the largest friction, $u_s$, is anti-rolling: the disc slides along the line and rotates in the opposite direction as it would for rolling, so its contact point slips twice as much as it would for pure sliding (velocity $\propto(1,0)$). For pure sliding, the magnitude of the friction will be roughly half this, about $(\lambda_s+\lambda_r)/2$. Note the force will not be aligned with the motion, but rather will have both a sliding component and a torque component. 

Returning to our original question -- is it more likely for the disc to roll or to slide? --  we see the answer for short, stiff DNA strands is unambiguously rolling, the friction for sliding being larger by a factor of about $\theta^{-2}$, where $\theta\sim (y-x)/\bar l\ll 1$ is small. 
What is the physical reason why rolling has the smallest friction? 
The main observation is that when the disc rolls, the contact point C is momentarily stationary, so any DNA tethers bound at this point will neither stretch nor bend, hence will incur no energetic penalty. 
The only term that makes the rolling friction non-zero comes from the bending energy associated with nearby DNA tethers, whose angles change quadratically with the amount of rolling but whose lengths change at higher order, leading to friction for this motion $\propto k_\theta \theta^2$.
When the disc slides, however, the angles of the DNA tethers bound at C change linearly with the amount of displacement, and the lengths change at quadratic order. The friction correspondingly contains a term $\propto k_{\theta}$ from the bending energy, and a term $\propto k_l \theta^2 \sim k_l\Delta l$ from the stretching energy.

We now see that the velcro analogy from the introduction does give intuition into the dynamics of DNA-coated surfaces, and helps to explain why rolling has the smallest friction. Although the physics of velcro is not identical to the physics of sticky ssDNA, 
the differences in forces required to move the system in different directions play an important role in both systems. 
Velcro requires a large force to slide but a smaller one to peel off, and similarly DNA-coated surfaces require larger forces instantaneously to slide along each other than to roll. Although the DNA is constantly and rapidly binding and unbinding, it still remains bound for long enough during a displacement to transmit the instantaneous forces to the opposite surface, and since these forces are strong because of the lengthscale separation between the DNA and the surface displacements, the difference in forces can become significant.

\section{Derivation of the coarse-grained dynamics: One tether}\label{sec:1tether}

In this section we derive the coarse-grained dynamics of the interval over long timescales. We start by considering a single tether because the derivation is easier to follow; the case of $N>1$ tethers is formally the same but technically more involved, and is dealt with in the ESI$^\dag$. Since there is only one tether we omit the subscript and call its length $l$ and its state $s$; recall $s=b$ if the tether is bound and $s=u$ if the tether is unbound. 
 
 The model outlined in section \ref{sec:1d} is a time-homogeneous Markov process $(v_t,l_t,s_t)$, whose dynamics are completely characterized by its generator $\Lgen$. Recall that, from the generator, the evolution of any statistic can be computed using the Kolmogorov backward equation \cite{Gardiner:2004bs}: given a function on the state space $g(v,l,s):\R^2\times \{u,b\} \to \R$, the statistic $f(v,l,s,t)\equiv \E^{v,l,s}g(v_t,l_t,s_t)$ evolves according to 
\begin{equation}\label{backward}
\partial_t f = \Lgen f, \qquad f(v,l,s,0) = g(v,l,s)\,,
\end{equation}
with appropriate boundary conditions. For our purposes it is sufficient that $|\nabla f|$ decay at infinity, which is equivalent to a no-flux condition at infinity for the corresponding Fokker-Planck equation.
Here $\E^{x}(\cdot)$ means the expectation given initial condition $x$. 

Our overall approach to obtaining the coarse-grained dynamics will be to consider the backward equation \eqref{backward} asymptotically in $\eps$, and to use singular perturbation theory to formally homogenize over the fast dynamics \cite{Pavliotis:2008wg}. 

We start by writing down the generator associated with the stochastic differential equations (SDEs) \eqref{tether_SDE}, \eqref{interval_SDE}, and the binding/unbinding dynamics. Since the state space is a mixture of continuous and discrete components, we must combine techniques for diffusion processes with techniques for Markov chains. We find it convenient to write the functions acted upon by the generator as a vector-valued function ${\bff(v,l,s,t)}=(f_u(v,l,t),f_b(v,l,t))^T$, such that the first component of the vector corresponds to setting $s=u$ and the second to $s=b$. The generator will then be a matrix-valued operator, and by examining the dynamical equations, we find

\begin{widetext}
\begin{equation}\label{Lsum}
\Lgen = 
\underbrace{\begin{pmatrix} -q_{on} & q_{on} \\ q_{off} & -q_{off}  \end{pmatrix}}_{\equiv Q} % Q
+ \underbrace{\begin{pmatrix} -\frac{k}{\gamma}l\partial_l + (\beta\gamma)^{-1}\partial^2_l & 0 \\ 0 & 0 \end{pmatrix}}_{\equiv U} % U
+ \underbrace{\begin{pmatrix} 0 & 0 \\ 0 & v\partial_l \end{pmatrix}}_{\equiv B} % B
+ \underbrace{\begin{pmatrix} 0 & 0 \\ 0 & -\frac{k}{m}l\partial_v \end{pmatrix}}_{\equiv V} \,.% V
\end{equation}
\end{widetext}
We have separated the generator into different components corresponding to different dynamical processes: $Q$ is the generator for the binding/unbinding dynamics, $U$ is the generator for the unbound tether dynamics, $B$ is the generator for the bound tether dynamics, $V$ is the generator for the interval's velocity dynamics. The full generator is a sum of the generators for each dynamical component as $\Lgen = Q+U+B+V$. 

Under the scalings given in \eqref{scalings}, we may verify that after nondimensionalizing the backward equation \eqref{backward}, the components of the generator scale as follows: $Q, U \sim O(\eps^{-2})$, $B,V\sim O(\eps^{-1})$. 
Therefore, if we separate out the small parameter explicitly, we may write the generator as  
\begin{equation}
 \label{eps_generator}
 \Lgen = \frac{1}{\eps^2} \left(\tilde Q + \tilde U\right) + \frac{1}{\eps} \left(\tilde V + \tilde M\right) \, .
\end{equation}
Here $\tilde Q = \eps^2 Q \sim O(1)$, and similarly for the other components. 

Let us assume that $\bff$ has an asymptotic expansion
\begin{equation}
 \label{eps_ansatz}
 \bff = \bff^{(0)} + \eps \bff^{(1)} + \eps^2 \bff^{(2)} + \ldots\,.
\end{equation}
We substitute this ansatz into the backward equation \eqref{backward} and solve at each power of $\eps$.

At order $O(\eps^{-2}$), we have
\begin{equation}
 \label{single_eps_m2}
 \left(Q+U\right)\bff^{(0)} = 0.
\end{equation}
To solve this for $\bff^{(0)}$, notice that the matrix $Q+U$ is row equivalent to the matrix 
\[
\begin{pmatrix}-1 & 1 \\ -\frac{k}{\gamma}l\partial_l + (\beta\gamma)^{-1}\partial^2_l  & 0 \end{pmatrix}\,.
\]
Therefore $\bff^{(0)}(v,l,t) = (1,1)^TC(v,l,t)$ where $\big(-\frac{k}{\gamma}l\partial_l + (\beta\gamma)^{-1}\partial^2_l \big)C = 0$. The only solution $C$ that satisfies the appropriate decay boundary conditions is a constant with respect to $l$, so the unique solution for $\bff^{(0)}$ (up to constants) is
\begin{equation}
 \bff^{(0)} = \begin{pmatrix}1 \\ 1\end{pmatrix} a(v,t)\,.
\end{equation}
Here $a(v,t)$ is some unknown function of $v$ and $t$, whose evolution will be determined after solving at higher order in $\eps$.

At order $O(\eps^{-1})$, we obtain
\begin{equation}
 \label{single_eps_m1}
 \left(Q+U\right)\bff^{(1)} = - (B + V) \bff^{(0)} = \begin{pmatrix}0 \\ 1\end{pmatrix} \frac{k}{m} l \D_{v} a\,.
\end{equation}
We can solve this for $\bff^{(1)}$ using the method of undetermined coefficients. 
Substituting the ansatz $\bff^{(1)} = (\alpha_u,\alpha_b)^T l$, where $\alpha_u$ and $\alpha_b$ are independent of $l$, into \eqref{single_eps_m1}, we can solve for $\alpha_u$ and $\alpha_b$ to obtain
\begin{equation}
 \label{single_eps_m1_soln}
 \bff^{(1)}  = \begin{pmatrix}f_u^{(1)} \\ f_b^{(1)}\end{pmatrix} = \begin{pmatrix}\alpha_u \\ \alpha_b\end{pmatrix} l
 = -\frac{1}{m q_{off}}  \begin{pmatrix} \gamma q_{on} \\ (k + \gamma q_{on})  \end{pmatrix} l \D_{v} a .
\end{equation}
In \eqref{single_eps_m1_soln}, we only consider the particular solution. Coefficients associated with the homogeneous solution will only be fixed at higher orders than we consider here.

It is here, in solving \eqref{single_eps_m1} to obtain \eqref{single_eps_m1_soln}, that we have used the assumption that the binding and unbinding rates, $\wt{q}_{on}$ and $\wt{q}_{off}$, are constant. Indeed, note from \eqref{single_eps_m1_soln} that $\alpha_b$ depends on $1/\wt{q}_{off}$. If the rate $\wt{q}_{off}$ depends on the tether length, $l$, then so does $\alpha_b$ and the expansion ansatz $\bff^{(1)} = (\alpha_u,\alpha_b)^T l$, with $\alpha_u$ and $\alpha_b$ constant in $l$, is not valid.

At order $O(\eps^0)$, 
\begin{equation}
 \label{single_eps_0}
 \left(Q+U\right)\bff^{(2)} = - (B + V) \bff^{(1)} + \D_t  \bff^{(0)} .
\end{equation}
By the Fredholm alternative, it is only possible to solve for $\bff^{(2)}$ if, for each $\bpi$ in the nullspace of $(Q+U)^*$, the adjoint of operator $Q+U$, we have 
\begin{equation}
\label{single_solvability}
\inner{\bpi,- (B + V) \bff^{(1)} + \D_t  \bff^{(0)}} = 0 \,.
\end{equation}
The inner product above is defined on vector-valued functions of $l$ to be $\inner{\mathbf{a},\mathbf{b}} = \int \mathbf{a}\cdot \mathbf{b} \: dl$. 
Physically, $\bpi$ represents the stationary distribution for the tethers when the variables characterizing the interval are frozen. The solvability condition \eqref{single_solvability} means that the slower dynamics associated to $\bff^{(1)}$,  must be orthogonal to $\bpi$, i.e. these dynamics don't have any component that can alter the tether's stationary distribution; if they did, then such alterations would build up on the fast timescale and would lead to exploding solutions on the slower timescale. 
Note that a similar solvability condition must hold for \eqref{single_eps_m1}, but one can check it is satisfied, as indeed it must be since we found a solution. 

After we solve for $\bpi$, we may substitute the known functional forms for $\bff^{(1)}$ and $\bff^{(0)}$, and perform the integral in \eqref{single_solvability} to obtain an evolution equation for $a(\wt{v},t)$.
One may verify that the only function in the nullspace of $(Q+U)^*$ (with decay conditions at $\infty$) is the one proportional to 
\begin{equation*}
%  \label{single_null}
 \bpi =  \begin{pmatrix}\frac{\wt{q}_{off}}{\wt{q}_{on}} \\ 1\end{pmatrix} e^{-\beta k l^2/2}\,.
\end{equation*}
Each of the terms appearing in the inner product in \eqref{single_solvability} can now be simplified:
\begin{align*}
\inner{\bpi, -B \bff^{(1)}} &= \frac{k+\wt{\gamma} \wt{q}_{on}}{\wt{m} \wt{q}_{off}}\ \wt{v} \D_{\wt{v}} a , \\
\inner{\bpi, -V \bff^{(1)}} &= -\frac{k+\wt{\gamma} \wt{q}_{on}}{\beta \wt{m}^2 \wt{q}_{off}}\ \D^2_{\wt{v}} a ,\\
\inner{\bpi, \D_t \bff^{(0)}} &= \left(1 + \frac{\wt{q}_{off}}{\wt{q}_{on}}\right) \ \D_t a \,.
\end{align*} 
Substituting the above relations into the solvability condition \eqref{single_solvability} and grouping terms gives the following equation:
\begin{equation}
 \label{solvability_reduce_1}
 \D_t a = \left( -\wt{\theta}_1 \wt{v} \D_{\wt{v}} + \frac{1}{2} \wt{\sigma}_1^2 \D^2_{\wt{v}} \right) a, 
\end{equation}
where 
\begin{equation}\label{theta_1_defn} 
 \wt{\theta}_1 = \frac{1}{m}\frac{1}{1+\frac{\wt{q}_{off}}{\wt{q}_{on}}}\frac{k+\gamma\wt{q}_{on}}{\wt{q}_{off}}\,,
 %\frac{\left( \frac{k+\wt{\gamma} \wt{q}_{on}}{\wt{m} \wt{q}_{off}} \right) }{ \left(\frac{\wt{q}_{off}}{\wt{q}_{on}}+1\right) }\,,  
 \qquad \text{and} \quad  
 \frac{1}{2}\wt{\sigma}^2_1 = \frac{1}{\wt{m}\beta} \wt{\theta}_1\,.
\end{equation}
Equation \eqref{solvability_reduce_1} is the backward equation for the process $V(t)$ that solves the SDE \eqref{SDEeffective}
%\begin{equation}\label{coarseSDE}
%m\dd{V}{t} = -\Gamma_{\rm DNA}V+ \sqrt{2\beta^{-1}\Gamma_{\rm DNA}}\eta(t)
%\end{equation}
with $\Gamma_{\rm DNA} = m\theta_1$ and $N=1$ (recall that $\wt{q}_{off}/\wt{q}_{on}=e^{-\beta e_0}$, from \eqref{binding_rates}.) 
Note there could be other SDEs with the same backward equation, but their solutions would all be weakly equivalent to $V(t)$.
Therefore the friction in the case $N=1$ has the formula \eqref{FrictionDNA} as claimed.

\section{Discussion \& Conclusion}\label{sec:conclusion}

We have constructed and analyzed a one-dimensional model for a colloidal particle moving on a DNA-coated surface. We showed that when the DNA evolves much faster than the particle, then on long timescales the particle evolves according to a Langevin equation with effective, DNA-induced friction coefficient $\Gamma_{\rm DNA}$. We gave an analytic expression for $\Gamma_{\rm DNA}$ in terms of the physical parameters of the DNA. Substituting values for these parameters characteristic of DNA-coated colloids with diameters $\approx1\mu\text{m}$ which can rearrange while bound, gave an estimate for the DNA-induced friction about 100 times larger than the hydrodynamic friction felt by such particles. 
We explored a mean-field extension of our model to two dimensions, by considering a disc moving along a line, and showed the friction when the disc rolls approaches zero as the DNA strands become stiffer and shorter, while the friction for the disc sliding remains comparatively large. 

If our predictions are correct, they not only imply that DNA-induced friction should be incorporated into numerical simulations of DNA-coated colloids, but they could also have significant implications for how DNA-coated colloids self-assemble. They imply that the dynamics of rearrangements are slow, orders of magnitude slower than the dynamics of particles interacting by depletion, for example, which are affected only by hydrodynamic friction, causing DNA-coated particles to stay longer in metastable states and possibly to never reach equilibrium. 
In fact, for small systems one might be able to make dynamical predictions using only DNA-induced friction, neglecting hydrodynamic interactions entirely; for larger systems we expect long-range hydrodynamic interactions to become equally or more important. If particles prefer to move in ways that minimize the amount of surface-surface rubbing, then dynamically they may behave like gears, and may get stuck for long times in more open arrangements where the number of contacts is less than the isostatic number needed to stabilize spheres mechanically. It has been observed in recent experiments that crystals of DNA-coated colloids like to grow as open, ordered structures, such as in a diamond lattice, transitioning only once they are big enough to more compact structures with more contacts and lower energy \cite{Wang:2017bd}; 
an intriguing possibility is whether this unusual growth is aided in part by the relative dynamics of the DNA-based interactions, such as a preference for rolling \cite{HolmesCerfon:2016ji}. (Note that reaction-limited kinetics, another possible effect of DNA-induced interactions, can also lead to low-coordination open structures  \cite{JanBachmann:2016dm}.) 

Because of these possible implications it would be extremely useful to test our predictions experimentally. A direct test could consider a DNA-coated colloid moving on a DNA-coated plane, and measure the diffusion coefficient as a function of the temperature, the length of the DNA strands, etc. If rotational degrees of freedom are accessible experimentally then one could also compare the diffusion coefficient for rolling, to that for sliding, to see if the difference is large, as predicted.

Our model was a drastic simplification of the full system and there are many omitted physical properties that would be interesting to study. Most critically, we hope to extend our model to higher dimensions, to more completely and rigorously study a disc moving on a surface, including vertical deviations from the surface, or eventually a sphere moving on a sphere. 
It is not yet clear whether such an extension would be possible with the techniques we have developed here -- it could be that, because the tethers are bound at fixed locations on the substrate, the effective friction will depend on precisely where the particle is relative to the fixed, discrete tethers, so the spacing between tethers may enter in the final model. Such graininess is unnecessary but to avoid it may require working with a density of tethers, which would evolve according to a stochastic \emph{partial} differential equation, requiring more sophisticated techniques to analyze and homogenize. 
It could also be that our techniques will be directly extendable to the higher-dimensional setting, if the friction comes from a superposition of forces from individual tethers, so that we may simply integrate the friction for one tether, with respect to the number density of tethers.  
It would be useful therefore to find a simpler way to derive the N-tether result, or to prove the heuristic formula \eqref{heuristic} in a more general setting. 
One important consideration in two dimensions is constructing a model for the energy which respects the geometry of the bodies on contact; a tether binding to a sloped or curved line, may not be as easily described as a tether binding to a flat surface (although see the detailed model in \cite{AngiolettiUberti:2016dd}.) 

Some additional effects that are worthwhile to investigate include: 
\begin{enumerate*}[(i)]
\item What is the effect of a tether binding to a tether, and not to a fixed spot on the surface? Experimentally both surfaces in contact are coated with sticky tethers, which all evolve and bind to each other. This could have the effect of ``smearing out'' the sharp separation in friction coefficients for different directions of motion, or, it could introduce a negligible perturbation to our results. The physical nature of the interaction can be even more complicated; in some systems the tethers are bound indirectly by linkers, making three bound polymers for each interaction \cite{Rogers:2015bv}. 
\item What is the effect of DNA strands binding non-independently of each other? When two DNA strands are bound, this reduces the possible partners for neighbouring DNA strands. Equilibrium studies show that neglecting such effects leads to overestimating the effective binding energy between colloids, even when the DNA is densely grafted \cite{AngiolettiUberti:2013iu}, suggesting that such neglect might also overestimate the effective friction, since friction increases with the overall binding energy. 
\item How does the effect of confinement, change the thermodynamics and dynamics of the DNA tethers from those predicted by the WLC model? Studies of the effective interactions in equilibrium have shown that the geometry of confinement can change the predicted pair interaction well depth by several $k_BT$ \cite{Dreyfus:2009gl,Rogers:2011et,Varilly:2012gla}, so there could be a similarly important effect on the dynamical properties. 
Relatedly, even the specific sequence of nucleotides can change the entropies and bending rigidity of ssDNA \cite{Goddard:2000ug}. 
\item What is the effect of a length-dependent unbinding rate, $q_{off}=q_{off}(l)$? 
A constant rate is reasonable for DNA unbinding, but may not be for other ligand-receptor systems. 
While the friction coefficient may no longer be possible to compute analytically, it may still be obtained through numerical integration. 
\item What if the binding/unbinding rates occur on a slow timescale, as they do in certain experimental systems? \cite{Parolini:2016ho,JanBachmann:2016dm,Zhang:2017kw}. It may be possible to systematically derive a formula for the effective friction and diffusivity, by considering a different scaling ansatz. 
\item What is the effect of spatial inhomogeneity of tethers? Early work has suggested this can sometimes cause the tethers to undergo subdiffusion \cite{Xu:2011gs}, a phenomenon we can't access with the formal homogenization techniques used here, but which may still be understood by analyzing the variation in diffusivity over larger spatial scales, or appealing to the literature on random walks in random environments \cite{statistics:ti}. 
\item How can one incorporate hydrodynamic interactions into the model, to predict a final friction coefficient that is some blend of the hydrodynamic friction and the DNA-induced friction? 
We discussed how a Langevin model is inappropriate for describing hydrodynamic interactions. It could be that one can incorporate DNA dynamics directly into an overdamped Langevin equation, or that one can start the analysis with a generalized Langevin equation, an exact model for hydrodynamics which includes  memory effects \cite{Bian:2016hp}. 
\end{enumerate*}

We hope that with these extensions, we may derive a quantitative prediction for the DNA-induced friction tensor in a wide variety of situations, which may be used as input to dynamical simulations or as the starting point for physically understanding out-of-equilibrium behaviour of DNA-coated colloids. Eventually, one might hope to control the friction, either to minimize it, or to design the most likely rearrangement pathways so particles may self-assemble into some desired, possibly nonequilibrium structure.

%((** conclusion? Notice that the bending energy is critical here to obtaining nonzero terms in the friction of different motions: because the geometry imposes that vertical tethers be displaced mainly orthogonally to their lengths,  stretching contributes negligibly compared to bending; contrast this to the one-dimensional setting where the tethers could only stretch parallel to their lengths. It would be worthwhile therefore to reevaluate the model for bending energy, incorporating more complex physics and geometry, since it seems so critical in two or more dimensions. ))

% If you have acknowledgments, this puts in the proper section head.
\begin{acknowledgments}
We would like to thank L. Mahadevan, Peter Kramer, and Aleks Donev for helpful discussions. 
 This work was supported primarily by the MRSEC Program of the National Science Foundation under Award Number DMR-1420073. 
 M.H.-C. was partially supported by Department of Energy Grant DE-SC0012296 and the Alfred P. Sloan foundation. 
\end{acknowledgments}

\begin{widetext}
% Specify following sections are appendices. Use \appendix* if there is only one appendix.
 \appendix
\section{Details of disc calculations}\label{sec:disccalcs}

This section contains auxiliary details of the calculations from section \ref{sec:2d}.

The first derivatives $\partial (l,\theta)/ \partial (x,\xi)$ are calculated by implicit differentiation from the relations \eqref{attach_horizontal}, \eqref{attach_vertical} to be
\begin{equation}
\frac{\partial l}{\partial x} = \sin \theta, \quad 
\frac{\partial l}{\partial \xi} = R\sin(\xi+\theta),\quad 
\frac{\partial \theta}{\partial x} = \frac{\cos\theta}{l},\quad 
\frac{\partial \theta}{\partial \xi} = \frac{R}{l}\cos(\xi+\theta)\,.
\end{equation}
The second derivatives $\partial^2(l,\theta) / \partial(x,\xi)^2$ are
\begin{gather}
\nonumber
\frac{\partial^2 l}{\partial x^2} = \frac{\cos^2\theta}{l}, \quad  
\frac{\partial^2 l}{\partial \xi^2} = \frac{R^2}{l}\cos(\xi+\theta)\left(\frac{l}{R}+\cos(\xi+\theta)\right), \quad  
\frac{\partial^2 l}{\partial x\partial \xi} = \frac{R}{l}\cos\theta\cos(\xi+\theta),   \\
\frac{\partial^2 \theta}{\partial x^2} = \frac{\sin 2\theta}{l^2}, \quad
\frac{\partial^2 \theta}{\partial \xi^2} = \frac{R^2}{l^2}\sin(\xi+\theta)\left( \frac{l}{R}+2\cos(\xi+\theta) \right), \quad
\frac{\partial^2 \theta}{\partial x\partial \xi} = \frac{-R}{l^2}\sin(\xi+2\theta) \,.
\end{gather} 

The second derivatives of energy, calculated by applying the chain rule to expression \eqref{Etether2d},  are
\begin{align*}
\partial_{xx} \Et &= k_l\Big(\frac{\partial l}{\partial x}\Big)^2 + k_{\theta}\Big(\frac{\partial \theta}{\partial x}\Big)^2 + k_l (l-\bar l)\frac{\partial^2 l}{\partial x^2} + k_\theta \theta \frac{\partial^2 \theta}{\partial x^2} \\
&= k_l + \frac{k_\theta}{l^2}(\cos^2\theta + \theta \sin 2\theta) \\
\partial_{\xi\xi}\Et &=  k_l\Big(\frac{\partial l}{\partial \xi}\Big)^2 + k_{\theta}\Big(\frac{\partial \theta}{\partial \xi}\Big)^2 + k_l (l-\bar l) \frac{\partial^2 l}{\partial \xi^2} + k_\theta \theta \frac{\partial^2 \theta}{\partial \xi^2} \\
&= k_l R^2\left(1+\frac{l}{R}\cos(\xi+\theta)\right)\\
& \qquad + \frac{k_\theta R^2}{l^2}\left( \cos^2(\xi+\theta) + \theta\sin(2(\xi+\theta)) + \frac{l}{R}\theta\sin(\xi+\theta)\right)\\
\partial_{x\xi} \Et &= k_l\Big(\frac{\partial l}{\partial x} \Big)\Big(\frac{\partial l}{\partial \xi}\Big) + k_\theta\Big(\frac{\partial \theta}{\partial x} \Big)\Big(\frac{\partial \theta}{\partial \xi}\Big)
 +k_l (l-\bar l) \frac{\partial^2 l}{\partial x\partial \xi} + k_\theta\theta\frac{\partial^2\theta}{\partial x\partial \xi}\\
  &= k_l R\cos\xi + \frac{k_\theta R}{l^2} \left(\cos\theta\cos(\xi+\theta) - \theta\sin(\xi+2\theta) \right)
\end{align*}

\section{Derivation of the coarse-grained dynamics: N tethers}\label{sec:Ntethers}

In this section we derive the coarse-grained dynamics of the full system of N tethers. 
 We use the same asymptotic procedure as in Section \ref{sec:1tether} for the case of one tether, only now the generator is more complicated. We start by explicitly showing the generator for $N=3$ in order to better illustrate its structure, and then outline the generator and derivation of the coarse-grained dynamics for $N$ tethers. 

For $N=3$ tethers, the set of possible states for the collection of tethers is $\{u,b\}^3$ so we need a vector of size  $2^3=8$ to represent the collection of states: 
\begin{equation} \label{3_tether_states}
\bff = (f_{uuu},f_{uub},f_{ubu},f_{buu},f_{ubb},f_{bub},f_{bbu},f_{bbb})^T.
\end{equation}
For example, the state where all tethers are unbound is identified with the component $f_{uuu}$ and the state where only the second tether is bound to the interval is identified with the component $f_{ubu}$. 

The full generator will still have the abstract decomposition $\Lgen = Q+U+B+V$ as in \eqref{Lsum}, but now the generator for each sub-process will be an $8\times8$ matrix of operators.  We now write down each generator in turn. 

The generator associated with the binding/unbinding dynamics is 
\begin{equation}
 \label{3_tether_binding_generator}
 Q^T = 
 \begin{pmatrix}
 -3\lambda & \nu & \nu & \nu & 0 & 0 & 0 & 0 \\
 \lambda & -2\lambda-\nu & 0 & 0 & \nu & 0 & 0 & 0 \\
 \lambda & 0 & -2\lambda-\nu & 0 & \nu & 0 & \nu & 0 \\
 \lambda & 0 & 0 & -2\lambda-\nu & 0 & \nu & \nu & 0 \\
 0 & \lambda & \lambda & 0 & -\lambda-2\nu & 0 & 0 & \nu \\
 0 & \lambda & 0 & \lambda & 0 & -\lambda-2\nu & 0 & \nu \\
 0 & 0 & \lambda & \lambda & 0 & 0 & -\lambda-2\nu & \nu \\
 0 & 0 & 0 & 0 & \lambda & \lambda & \lambda & -3\nu
 \end{pmatrix}  ,
\end{equation}
where, for display purposes, we have defined $\lambda \equiv q_{on}$, and $\nu \equiv q_{off}$.

The generator for the evolution of the unbound tether lengths, 
following the ordering of the states given in \eqref{3_tether_states}, is the diagonal matrix
\begin{equation*}
%  \label{3_tether_unbound_generator}
 U = \diag \left( \sum_{j=1,2,3}\Lgen^{OU}_j, \sum_{j=1,2}\Lgen^{OU}_j, \sum_{j=1,3}\Lgen^{OU}_j, \sum_{j=2,3}\Lgen^{OU}_j, \Lgen^{OU}_1, \Lgen^{OU}_2, \Lgen^{OU}_3, 0 \right) .
\end{equation*}
where
\begin{equation}
 \label{j_OU_generator}
 \Lgen_j^{OU} = -k \gamma^{-1} l_j \D_{l_j} + \gamma^{-1} \beta^{-1} \D_{l_j}^2 \,.
\end{equation}
For example, the element $U_{33} = \sum_{j=1,3}\Lgen^{OU}_j$ acts on $f_{ubu}$, corresponding to the state for which only tethers $j=1$ and $3$ are unbound.

The generator for the evolution of bound tethers is the diagonal matrix
\begin{equation*}
%  \label{3_tether_bound_generator}
 B = \diag \left( 0, \Lgen_3^{bd}, \Lgen_2^{bd}, \Lgen_1^{bd}, \sum_{j=2,3}\Lgen_j^{bd}, \sum_{j=1,3}\Lgen_j^{bd}, \sum_{j=1,2}\Lgen_j^{bd}, \sum_{j=1,2,3}\Lgen_j^{bd}  \right) 
\end{equation*}
where
\begin{equation}
 \label{j_bound_generator}
 \Lgen_j^{bd} = v \D_{l_j} \,.
\end{equation}
For example, the element $B_{33} = \Lgen_2^{bd}$ acts on $f_{ubu}$, corresponding to the state for which only the $j=2$ tether is bound.

The generator for the interval velocity dynamics for the $3$ tether system is the diagonal matrix
\begin{equation*}
%  \label{3_tether_interval_generator}
V = \diag\left( 0, \Lgen_3^{int}, \Lgen_2^{int}, \Lgen_1^{int}, \sum_{j=2,3}\Lgen_j^{int}, \sum_{j=1,3}\Lgen_j^{int}, \sum_{j=1,2}\Lgen_j^{int}, \sum_{j=1,2,3}\Lgen_j^{int}\right)
\end{equation*}
where
\begin{equation}
\label{j_interval_generator}
 \Lgen_j^{int}  = - \frac{k}{m} l_j \D_v \,. 
\end{equation}

\medskip

Now consider the generator for the $N$ tether system, which is a natural generalization of the generator for the $3$ tether system. 
For general $N\geq 1$, we need a vector of length $2^N$ to represent the possible states of the system:
\begin{equation}
 \label{N_tether_states}
\bff = (f_{u \ldots u},f_{u \ldots ub},f_{u \ldots ubu},f_{u \ldots ubuu},\ldots,f_{b \ldots bub},f_{b \ldots bu},f_{b \ldots b})^T.
\end{equation}
Suppose the states are labelled $1,2,\ldots,2^N$. Let $b(i),u(i)$ be the set of tethers which are bound/unbound in state $i$ respectively. For example, $b(3) = \{2^N-1\}$ and $u(3) = \{1,2,\ldots,2^N-2,2^N\}$. 
Let $|b(i)|,|u(i)|$ be the number of bound/unbound tethers. Clearly $|b(i)| +|u(i)|=N$. 

It will be convenient to define a matrix $S\in \{0,1\}^{2^N\times N}$ to be the matrix whose rows (corresponding to different states) are a set of flags indicating whether each tether (the columns) is bound (1) or unbound (0) in each state, i.e. $S_{ij}=1$ if tether $j\in b(i)$, $S_{ij} = 0$ if tether $j\in u(i)$. 
The matrix which flags unbound tethers is $\underline{1}-S$, where $\underline{1}$ is the $2^N\times N$-dimensional matrix whose entries are all 1. 

The generator $Q$ for the binding and unbinding process has components
\begin{equation}\label{Q_N_defn}
Q_{ij} = \left\{ \begin{array}{cl}
\lambda & \text{if } |b(j)| = |b(i)|+1\\
\nu & \text{if }|u(j)| = |u(i)|+1\\
-\lambda |u(i)| - \nu |b(i)|& \text{if } i=j
\end{array}\right.
\end{equation}

The $N$ tether generalizations of the $U$, $B$, and $V$ matrices are 
\begin{align}
 U &= \diag \left( \sum_{j=1,\ldots,N}\Lgen^{OU}_j, \sum_{j=1,\ldots,N-1}\Lgen^{OU}_j ,\ldots, \sum_{j=N,N-1}\Lgen^{OU}_j, \Lgen^{OU}_1, \ldots, \Lgen^{OU}_{N-1}, \Lgen^{OU}_N, 0 \right) \nonumber \\
 &= \diag\left((\underline{1}-S)\Lgen^{OU} \right),   \label{U_N_defn}  \\
 B &= \diag \left( 0, \Lgen_N^{bd}, \ldots, \Lgen_1^{bd}, \sum_{j=N,N-1}\Lgen_j^{bd}, \ldots, \sum_{j=1,\ldots,N-1}\Lgen_j^{bd}, \sum_{j=1,\ldots,N}\Lgen_j^{bd}  \right) \nonumber \\
 &= \diag\left(S\Lgen^{bd} \right), \label{B_N_defn} \\
V &= \diag \left( 0, \Lgen_N^{int}, \ldots, \Lgen_1^{int}, \sum_{j=N,N-1}\Lgen_j^{int}, \ldots, \sum_{j=1,\ldots,N-1}\Lgen_j^{int}, \sum_{j=1,\ldots,N}\Lgen_j^{int}  \right), \nonumber\\
&= \diag\left(S\Lgen^{int} \right)  \label{V_N_defn}   
\end{align}
The operators $\Lgen^{OU}_j$, $\Lgen_j^{bd}$, and $\Lgen_j^{int}$ are defined in \eqref{j_OU_generator}, \eqref{j_bound_generator} and \eqref{j_interval_generator}, respectively, 
and $\Lgen^{OU} = (\Lgen^{OU}_1,\ldots,\Lgen^{OU}_N)^T$, and similarly for the other operators. 

% The elements of the matrices $U$, $B$, $M$ and $H$ are the elements of the power set of $\{\Lgen_j^{x}\}_{j=1,\ldots,N}$, where we identify subsets $\{\Lgen_m^{x},\Lgen_n^{x}\}$ as $\sum_{j=m,n}\Lgen_j^{x}$ and the null set as $0$. 
%The relative ordering of the diagonal elements in the matrices reflects the ordering introduced in \eqref{N_tether_states}.

With the generator in hand we may proceed with the asymptotic analysis. 
We substitute the ansatz \eqref{eps_ansatz} into the backward equation \eqref{backward} and equate equal powers of $\eps$ to obtain a hierarchy of equations governing $\bff^{(i)}(\bfl,\wt{v},t)$.
At order $O(\eps^{-2}$), we have
\begin{equation}
 \label{N_eps_m2}
 \left(Q+U\right)\bff^{(0)} = 0.
\end{equation}
Equation \eqref{N_eps_m2} possesses only constant in $\bfl$ solutions of the form:
\begin{equation}
 \label{N_eps_m2_soln}
 \bff^{(0)} = \left(1,1,\ldots,1\right)^T a(\wt{v},t) .
\end{equation}
At order $O(\eps^{-1})$, we have
\begin{equation}
 \label{N_eps_m1}
 \left(Q+U\right)\bff^{(1)} = - (B + V) \bff^{(0)} = -V \bff^{(0)} = \frac{k}{m}\partial_v a \: S\bfl\,.
\end{equation}
We have used that $B \bff^{(0)}=0$ because $\bff^{(0)}$ is independent of $l_j$, for $j=1,\ldots,N$; see \eqref{B_N_defn} and \eqref{j_bound_generator}.

We claim the solution is 
\begin{equation}\label{f1N}
\bff^{(1)} = -\frac{k}{m}\partial_v a\left( \frac{\lambda}{\nu}\frac{\gamma}{k}  \underline{1} + \frac{1}{\nu}S \right)\bfl\,,
\end{equation}
To show this, it is sufficient to show that
\[
 \left(Q+U\right)\left( \frac{\lambda}{\nu}\frac{\gamma}{k} \underline{1} + \frac{1}{\nu}S \right)\bfl = -S\bfl\,.
\]
We calculate each of the four terms in the product on the left-hand side in turn. 

\begin{enumerate}
\item We have $Q\underline{1} = 0$, the zero matrix, since the sum of each row of $Q$ is 0. 

\item We also have that $US\bfl = 0$, since the $i$th entry is $((\underline{1}-S)\Lgen^{OU})_i\cdot (S\bfl)_i$, and $((\underline{1}-S)\Lgen^{OU})_i$ only contains operators acting on tethers in set $u(i)$, while $(S\bfl)_i$ only contains tether lengths from set $b(i)$. 

\item We have that $U\underline{1}\bfl = -\frac{k}{\gamma}(\underline{1}-S)\bfl$, since the vector $\underline{1}\bfl$ has components identically equal to $l_1+\cdots+l_N$, so all operators in each diagonal element have an effect, and $(U\underline{1}\bfl)_i=\sum_{j\in u(i)}\Lgen^{OU}_j l_j$. 

\item The remaining term to evaluate is $QS$. Consider component $(i,k)$: 
\[
(QS)_{ik} = \sum_j Q_{ij}S_{jk}\,.
\]
Recall that $S_{jk} = 1$ if $k\in b(j)$, $S_{jk}=0$ if $k\in u(j)$. 

Suppose $k\in b(i)$. Then term $Q_{ij}S_{jk} = \lambda$ iff $|b(j)|=|b(i)|+1$ and $k\in b(j)$. The number of such states, is the number of ways of adding a bound tether, since $k\in b(i)$ so we automatically have $k\in b(j)$. The contribution to the sum is $\lambda |u(i)|$. 
Now consider the number of $j$s such that $Q_{ij}S_{jk} = \nu$. We must have $|u(j)| = |u(i)| + 1$, and $k\in b(j)$. Without this last condition on $k$, we would have $|b(i)|$ such terms, the number of tethers that can be flipped from bound to unbound, however one of these flips is $k$ itself, which would make $S_{jk}=0$. Therefore the contribution to the sum is $\nu (|b(i)|-1)$. 
There is also a contribution from the diagonal, $Q_{ii}S_{ik} = -\nu|b(i)|-\lambda|u(i)|$. 
Putting this all together shows that 
\[
k\in b(i) \quad\Rightarrow\quad (QS)_{ik} = \lambda|u(i)| + \nu (|b(i)|-1) -\nu|b(i)|-\lambda|u(i)| = -\nu \,.
\]

Now suppose $k\in u(i)$. The number of $j$s that contribute a $\lambda$ equals 1, since to contribute we must have $k\in b(j)$, and so only $k$ can be flipped. The number of $j$s that contribute $\nu$ is 0, since a tether unbinds to go from $i\to j$ so $k\in u(j)$. 
Therefore
\[
k\in u(i) \quad\Rightarrow\quad (QS)_{ik} = \lambda \,.
\]

Putting all $k$s together shows that 
\[
QS = -\nu S + \lambda(\underline{1}-S)\,.
\]
\end{enumerate}

Now we put all the four terms together to find
\begin{align*}
 \left(Q+U\right)\left( \frac{\lambda}{\nu}\frac{\gamma}{k} \underline{1} + \frac{1}{\nu}S \right)\bfl  
 &= \left( -S + \frac{\lambda}{\nu}(\underline{1}-S)\right) \bfl - \frac{\lambda}{\nu}(\underline{1}-S)\bfl 
 = -S\bfl\,,
\end{align*}
so \eqref{f1N} holds, as claimed. 

Now we consider the $O(\eps^0)$ equation: 
\begin{equation}
 \label{N_eps_0}
 \left(Q+U\right)\bff^{(2)} = - (B + V) \bff^{(1)} + \D_t \bff^{(0)} .
\end{equation}
Solvability of \eqref{N_eps_0} requires that for each $\pi$ in the nullspace of $(Q+U)^*$, we have
\begin{equation}
\label{N_solvability}
\inner{\bpi,- (B + V) \bff^{(1)} + \D_t \bff^{(0)}} = 0.
\end{equation}
One can verify (in a similar way to the calculations in section \ref{sec:stationary}, see e.g. Eqs \eqref{pivl}, \eqref{pis}) that the nullspace of $(Q+U)^*$ is spanned by the vector 
\begin{align}
 \bpi &=  \left(\left(\frac{\nu}{\lambda}\right)^N,
 \underbrace{\left(\frac{\nu}{\lambda}\right)^{N-1},\ldots,\left(\frac{\nu}{\lambda}\right)^{N-1}}_{N \text{ elements}},
 \underbrace{\left(\frac{\nu}{\lambda}\right)^{N-2}, \ldots, \left(\frac{\nu}{\lambda}\right)^{N-2}}_{{N \choose 2} \text{ elements}}, \ldots, 1\right)^T 
 %& \qquad \qquad \qquad 
 \times e^{-\beta k \sum_{j=1}^N l_j^2/2}\,.\label{N_null}
\end{align}
We now compute each of the terms in the inner product. 

Letting $\mathbf b = (|b(1)|,\ldots,|b(2^N)|)^T$, we have
\begin{align*}
\inner{\bpi,-B\bff^{(1)}} &= \inner{\bpi,
-\frac{k}{m}\diag(S\Lgen^{bd})\left[\partial_v a \left( \frac{\lambda}{\nu}\frac{\gamma}{k}\underline{1} + \frac{1}{\nu}S \right)\bfl\right]
}\\
&= \left(\frac{\gamma\lambda+k}{m\nu}\right) v\partial_v a
\inner{\bpi, \mathbf b}\\
&= Z\: \left(\frac{\gamma\lambda+k}{m\nu}\right) v\partial_v a \:
\sum_{k=0}^N k{N \choose k} \left(\frac{\nu}{\lambda}\right)^{N-k}\\
&= Z\: \left(\frac{\gamma\lambda+k}{m\nu}\right) v\partial_v a \:
\sum_{k=1}^N N{N-1\choose k-1} \left(\frac{\nu}{\lambda}\right)^{N-k}\\
&= Z\:\left(\frac{\gamma\lambda+k}{m\nu}\right) v\partial_v a \:N\: 
\sum_{j=0}^{N-1} {N-1\choose j} \left(\frac{\nu}{\lambda}\right)^{N-1-j}\\
&= Z\:\left(\frac{\gamma\lambda+k}{m\nu}\right) v\partial_v a \: N\: 
\left( 1 + \frac{\nu}{\lambda}\right)^{N-1}
\end{align*}
where $Z = \int_{\R^N} e^{-\beta k \sum_{j=1}^N l_j^2/2}d\bfl$. \\

For the second term, we use that 
\[
\int_{\R^N} Z^{-1} l_m l_n e^{-\beta k \sum_{i=1}^N l_i^2/2} d\bfl= \frac{1}{k\beta} \delta_{m,n} , 
\]
and calculate the second inner product to be, substituting the calculation of $\inner{\bpi,\mathbf b}$ from above, 
\begin{align*}
\inner{\bpi,-V\bff^{(1)}} &= \inner{\bpi,
\frac{k}{m}\diag(S\Lgen^{int})\left[\partial_v a \left( \frac{\lambda}{\nu}\frac{\gamma}{k}\underline{1} + \frac{1}{\nu}S \right)\bfl\right]
}\\
&= -Z\:\frac{k}{m}\left(\frac{\gamma\lambda+k}{m\nu}\right) \partial^2_v a\:\frac{1}{\beta k}
\inner{\bpi,\mathbf b}\\
&= -Z\:\left(\frac{\gamma\lambda+k}{\beta m^2\nu}\right) \partial^2_v a\:
N\: \left( 1 + \frac{\nu}{\lambda}\right)^{N-1}
\end{align*}

The term involving $\bff^{(0)}$ 
in \eqref{N_solvability} are straightforward to simplify. We have
\begin{align*}
\inner{\bpi, \D_t \bff^{(0)}}&= \inner{\bpi,(1,1,\ldots,1)^T}\partial_t a\\
&= Z\:\sum_{k=0}^N{N \choose k}\left(\frac{\nu}{\lambda}\right)^{N-k} \\
&= Z\:\left( 1 + \frac{\nu}{\lambda}\right)^N\partial_t a 
\end{align*}

Substituting the above relations into the solvability condition \eqref{N_solvability} and grouping terms, \eqref{N_solvability} reduces to:
\begin{equation}\partial_t a = 
\frac{N\left(\frac{\gamma\lambda+k}{m\nu}\right) }
{1 + \frac{\nu}{\lambda}}\: v\partial_v a
+
\frac{1}{\beta m}\frac{N\left(\frac{\gamma\lambda+k}{m\nu}\right)}
{1 + \frac{\nu}{\lambda}} \: \partial^2_v a\,.
\end{equation}
This is the backward equation for the process that solves \eqref{SDEeffecive}, as claimed (recall $\lambda = q_{on}$, $\nu=q_{off}$, $\lambda/\nu = e^{\beta e_0}$.)

\section{Stationary distribution for the N tether dynamics}\label{sec:stationary}

In this section we verify that the stationary distribution for the $N$ tether dynamics is the Boltzmann distribution, $\pi \propto e^{-\beta E}$, where $Z$ is a normalizing constant and $E$ is the energy, given in \eqref{E}. 
Specifically, 
\begin{equation}\label{pi}
\pi(\bfl,\bfs,v) = Z^{-1} e^{-\frac{\beta mv^2}{2}} \prod_{j=1}^N e^{-\frac{\beta kl_j^2}{2}} \left( e^{\beta e_0 }  \delta_{s_j,b}
+ \delta_{s_j,u}  \right)\,.
\end{equation}
The normalization constant $Z$ is chosen to ensure that $\pi$ is a probability measure. 
From this formula, one can verify by direct integration that the probabilities a tether is bound or unbound at any length in equilibrium are those given in \eqref{pbpu}. 

It will be convenient to write $\pi$ in the following form: 
\begin{equation}
\pi = Z^{-1} \pi_{v,l}\:\bm{ \pi_s}
\end{equation}
where
\begin{equation}\label{pivl}
\pi_{v,l} = e^{-\frac{\beta mv^2}{2}}e^{-\frac{\beta k}{2} \sum_{j=1}^N l_j^2}
\end{equation}
and
\begin{equation}\label{pis}
\bm{\pi_s}  = \left(\left(\frac{\nu}{\lambda}\right)^N,
 \underbrace{\left(\frac{\nu}{\lambda}\right)^{N-1},\ldots,\left(\frac{\nu}{\lambda}\right)^{N-1}}_{N \text{ elements}},
 \underbrace{\left(\frac{\nu}{\lambda}\right)^{N-2}, \ldots, \left(\frac{\nu}{\lambda}\right)^{N-2}}_{{N \choose 2} \text{ elements}}, \ldots, 1\right)^T \,.
\end{equation}
The ordering of states is the same as that in \eqref{N_tether_states}. 

The generator is $\Lgen = Q+U+B+V$ with $Q,U,B,V$ defined in \eqref{Q_N_defn}, \eqref{U_N_defn}, \eqref{B_N_defn}, \eqref{V_N_defn} respectively. We must show that 
\[
\Lgen^* \pi = (Q+U+B+V)^*\pi = 0\,,
\]
where $^*$ denotes the formal adjoint. 

First we show that $(B+V)^*\pi_{v,l}\:\bm{c} = 0$, where $\bm c$ is any vector with the right dimensions which doesn't depend on $v,\bfl$. We have that
\[
(\Lgen^{bd}_j)^* \pi_{v,l} = -\partial_{l_j}(v\:\pi_{v,l}) = \frac{\beta kvl_j}{2}\pi_{v,l}
\]
and 
\[
(\Lgen^{int}_j)^* \pi_{v,l} = \frac{k}{m}\partial_v(l_j\: \pi_{v,l}) = -\frac{\beta kvl_j}{2}\pi_{v,l}\,.
\]
Therefore $(\Lgen^{bd}_j)^*\pi_{v,l}+(\Lgen^{int}_j)^*\pi_{v,l} = 0$, so $((B+V)^*\pi_{v,l}\bm c)_i = \bm c_i\sum_{j\in b(i)}\left((\Lgen^{bd}_j)^*\pi_{v,l}+(\Lgen^{int}_j)^*\pi_{v,l}\right)=0$, so the result follows.

Next we show that $U^*\pi_{v,l} = 0$. We have that
\[
(\Lgen^{OU}_j)^* \pi_{v,l} = \frac{k}{\gamma}\partial_{l_j}(l_j\pi_{v,l}) + \frac{\beta^{-1}}{\gamma}\partial^2_{l_j}\pi_{v,l} = 0\,,
\]
and therefore any sum $\sum_{j\in u(i)}(\Lgen^{OU}_j)^* \pi_{v,l} =0$. 

Finally we show that $Q^*\bm{\pi_s} = 0$. We have
\begin{align*}
(Q^*\bm{\pi_s})_i &= \sum_j Q_{ij}(\bm{\pi_s})_i \\
&= |b(j)|\lambda \left(\frac{\lambda}{\nu}\right)^{|u(j)|-1} + |u(j)|\nu\left(\frac{\lambda}{\nu}\right)^{|u(j)|+1} - (\lambda|u(j)| + \nu|b(j)|) \left(\frac{\lambda}{\nu}\right)^{|u(j)|}\\ &= 0
\end{align*}
Combining these calculations shows the desired result.

\end{widetext}

\bibliography{1D_colloid_diffusion,dnarefs}

\end{document}